\documentclass[aps,prd,twocolumn,superscriptaddress,floatfix,showpacs]{revtex4-1}
\usepackage{amsmath, amsfonts, mathrsfs}
\usepackage{graphicx}

\newcommand{\mnras}{Mon.\ Not.\ R.\ Astro.\ Soc.}

\newcommand{\aap}{A \& A}

\newcommand \beq {\begin{equation}}
\newcommand \eeq {\end{equation}}
\newcommand \beqn {\begin{eqnarray}}
\newcommand \eeqn {\end{eqnarray}}
\newcommand \apjl {ApJL}
\newcommand \apjs {Astrophys. J. Suppl.}

\begin{document}
\title{Accretion Disks Around Binary Black Holes: 
A Simple GR-Hybrid Evolution Model}
\author{Stuart L. Shapiro}
\altaffiliation{Also Department of Astronomy and NCSA, University of
  Illinois at Urbana-Champaign, Urbana, IL 61801}
\affiliation{Department of Physics, University of Illinois at
  Urbana-Champaign, Urbana, IL 61801}

\begin{abstract}
We consider a geometrically thin, 
Keplerian disk in the orbital plane of a binary black hole (BHBH)
consisting of a spinning primary and low-mass secondary (mass ratio
$q \lesssim 1$). To account for the principle effects of 
general relativity (GR), we propose a modification of the 
standard Newtonian evolution equation for the (orbit-averaged) 
time-varying disk surface density.  In our modified equation the 
viscous torque in the disk is treated in full GR,
while the tidal torque is handled in the Newtonian limit.
Our GR-hybrid treatment is reasonable 
because the tidal torque is concentrated near the orbital radius 
of the secondary and 
is most important prior to binary-disk decoupling, when the orbital
separation is large and resides in the weak-field regime. The
tidal torque on the disk diminishes during late merger and vanishes altogether
following merger. By contrast, the viscous torque drives the flow into the 
strong-field region and onto the primary
during all epochs. Following binary coalescence, the viscous torque 
alone governs the time-dependent accretion onto the remnant, as well as 
the temporal behavior, strength and
spectrum of the aftermath electromagnetic radiation from the disk. 
We solve our GR-hybrid equation for a representative BHBH-disk 
system, identify several observable EM signatures of the merger, 
and compare results obtained for the gas and EM radiation
with those found with the Newtonian prescription.

\end{abstract}
\maketitle

\section{Introduction}

Binary black hole (BHBH) mergers are likely to occur in regions immersed
in gas, and the capture and accretion of the gas by the binary may result in 
appreciable electromagnetic radiation.
There exists the realistic 
possibility of detecting electromagnetic ``precursor''
radiation prior to the merger and before the maximum gravitational wave 
(GW) emission from a BHBH merger~\cite{ArmN02,ChaSMQ10}. 
Then, following the detection of gravitational
waves, observing electromagnetic ``afterglow'' 
radiation could provide further confirmation of the 
coalescence~\cite{MilP05, RosLAPK09, SchK08, CorHM09, OneMBRS09, Sha10, 
TanM10}. Such electromagnetic radiation can also serve as a useful probe
of the gas in galaxy cores or in other regions where mergers take place,
as well as a diagnostic of the physics of black hole accretion. 
This diagnostic may be particularly revealing once 
the masses and spins of the merging companions and BH remnant are
determined from the GW signal.

In this paper we focus on a geometrically thin, Keplerian disk orbiting in the
plane of a spinning BH with a low-mass companion.
To follow the orbit-averaged, secular evolution of such a BHBH-disk system, 
a simplified, vertically integrated, $1+1$ - dimensional Newtonian model 
equation [Eq.~(\ref{eq:sigevol1}) below] has been adopted in many
previous studies (see, e.g.,~\cite{ArmN02, LodNKP09, ChaSMQ10} 
and references therein). 
We have demonstrated how the steady-state solution to this equation can
be used to determine the disk structure and electromagnetic radiation 
spectrum during the long inspiral epoch prior to binary-disk decoupling, during
which a quasistationary treatment is applicable~\cite{LiuS10}. 
To illustrate this approach we solved the steady-state equation for
representative BHBH-disk systems at decoupling, employing simplified 
prescriptions for the required viscosity $\nu(r)$ and disk scale-height 
$h(r)$ profiles. Our steady-state approach was extended 
in~\cite{KocHL12}, where the Shakura-Sunyaev~\cite{ShaS73}
``one-zone" prescription for radiation transport was adopted in conjunction 
with a ``$\beta$''-disk viscosity law to obtain these profiles 
self-consistently (see also~\cite{Raf12}).

Here we provide an alternative evolution equation that better approximates the
strong-field, relativistic nature of the circumbinary disk. This equation 
treats the viscous torque in full GR for gas flow in a thin Keplerian disk. 
The disk, which is not self-gravitating for densities of interest here, evolves
in the background spacetime determined by the more massive primary, 
assumed to be a (quasi-)stationary Kerr black hole. The tidal torque, 
arising from the presence of the low-mass secondary, is
handled in the Newtonian limit. The later approximation is reasonable 
since the tidal torque is strongly peaked in and just outside a narrow gap
in the disk centered on the orbit of the secondary. This torque
plays its most important role prior to binary-disk decoupling, when the
binary separation is large and lies outside the strong-field region of the
primary. Moreover, the tidal torque disappears altogether following merger.
By contrast, the viscous torque drives gas into the strong-field region
during all epochs, including the post-merger phase.
A GR treatment of the viscous-driven accretion onto the primary and the 
post-merger remnant is particularly important for making predictions of 
any observable, `precursor' and `aftermath' electromagnetic radiation that
may accompany the GW burst.

A fully reliable description of the accretion flow and associated
radiation really requires a
radiation magnetohydrodynamics (MHD) simulation in full general
relativity in the  $3+1$-dimensional, dynamical spacetime of the 
merging BHBH binary. Newtonian hydrodynamic simulations 
incorporating some of the relevant physics have been performed in 
various dimensions and levels of approximation 
(see, e.g.,~\cite{MacM08, CorHM09, RosLAPK09, OneMBRS09, ShiKLH12, DorHM12},
while GR simulations are in their preliminary stages
(e.g.,~\cite{MegAFHLLMN09,AndLMN10, FarLS10, BodHBLS10, MostaPRLYP10, 
PalenzuelaGLL10, ZanottiRDP10, FarLS11, BodBHHLS12}). 
Only recently have the
first relativistic MHD simulations of a BHBH-disk system been performed: 
Noble et al.~\cite{NobMNKCZY12} adopt post-Newtonian gravitation to 
perform simulations of an equal-mass system, 
excising the region inside the binary orbit, 
while Farris et al.~\cite{FarGPES12} summarize simulations in full GR that
cover the complete spatial domain, including the black holes.
Both of these relativistic MHD 
simulations deal primarily with geometrically thick (i.e. warm) disks 
in which an ``effective viscosity'' is provided by magnetic fields driven 
turbulent by the magnetorotational instability (MRI).

The model discussed here is mainly relevant for geometrically thin (i.e. cool) 
circumbinary disks. Although based on a simplified orbit-average description,
it should delineate many of the qualitative features characterizing the 
evolution of BHBH-thin disk systems.
Also, the model may be useful for selecting input parameters 
and identifying scaling behavior for future, more detailed numerical 
simulations.  In addition, the resulting solutions can provide approximate 
initial disk profiles for such simulations. It is in this spirit and toward 
these purposes that we propose the adoption of our simple GR-hybrid equation.
We hope that it provides a starting point for improved GR modeling along
these lines.

In Section II we review the Newtonian binary-disk model and the 
required elements that enter the secular evolution equation. We also 
summarize how the resulting accretion rate onto the primary and the
local electromagnetic flux and total luminosity from the disk can be 
calculated. Simplifications that arise when describing the pre-decoupling and
post-merger epochs are summarized. In Section III we 
present the GR-hybrid model and retrace our previous discussion, now adapting
it to the GR-hybrid equation. In Section IV we provide a numerical 
example by solving the equations for an illustrative BHBH-disk system. 
We begin our integrations prior to binary-disk decoupling and proceed through
inspiral and merger,
comparing the Newtonian and GR-hybrid solutions. In Section V we outline 
future work that will improve the model.
We adopt geometrized units and set $G=1=c$ throughout.

\section{The Newtonian Evolution Equations}
\subsection{Disk Evolution}

For reference and comparison we write down the standard Newtonian
evolution equation for the surface density, $\Sigma(t,r)$,
\begin{equation}
\label{eq:sigevol1}
\frac{\partial \Sigma}{\partial t} = - \frac{1}{2 \pi r} 
\frac{\partial}{\partial r} 
\left[
\left( \frac{\partial (r^2 \Omega)}{\partial r} \right)^{-1}
\frac{\partial G}{\partial r} 
\right].
\end{equation}
Here $G \equiv -T_{\rm vis} + T_{\rm tid}$ is the total torque, $T_{\rm vis}$ is the
viscous torque, $T_{\rm tid}$ is the tidal torque on the disk from the presence of
the secondary,
and $\Omega = \Omega_K = (M/r^3)^{1/2}$ is the Keplerian orbital frequency
about the primary, centered at $r=0$. The mass of the primary is $M$ and
the secondary $qM$, where $q \ll 1$.
The viscous torque density is
given by the standard equation~\cite{Pri81,FraKR02,ArmN02,ChaSMQ10}
\begin{equation}
\frac{\partial T_{\rm vis}}{\partial r} = 
-\frac{\partial}{\partial r} \left( 2 \pi r^3 \nu \Sigma 
\frac{\partial \Omega}{\partial r} \right).
\label{eq:dr_Tvis}
\end{equation}
We approximate the (orbit-averaged) tidal torque density
by using the expression adopted by Armitage and Natarajan~\cite{ArmN02}
\beq
\label{eq:AN}
\frac{\partial T_{\rm tid}}{\partial r}  = 2 \pi \Lambda  \Sigma r \,
\eeq
where $\Lambda(r,a)$ is given by
\beq
\label{eq:torqtid}
\Lambda = \left \{ \begin{array}{ll}
-\left (f q^2 M/2r \right) \left( r /\Delta_p \right)^4,   & r < a \\
+\left (f q^2 M/2r \right) \left( a /\Delta_p \right)^4,  & r > a \\
     \end{array} \right. \ .
\eeq

In Eq.~(\ref{eq:torqtid}) $f$ is a dimensionless
normalization factor
and $\Delta_p$ is given by $\Delta_p = {\rm max}(|r-a|, h)$, and
$a(t)$ is the orbital radius of the secondary.
Calibrating the above expression for the tidal field
against high-resolution, hydrodynamical
simulations in two-dimensions for a low-mass, black hole secondary
interacting with an outer
accretion disk, Armitage and Natarajan find that the
value $f \approx 0.01$ best fits the
simulation results. Equations~(\ref{eq:AN}) and (\ref{eq:torqtid})
furnish a reasonable analytic approximation
to the results obtained from summing over the pointlike
contributions from the Lindblad resonances in the disk~\cite{GolT80,HouW84}.
(Similar, but slightly different, forms for the tidal
torque also have been used in the literature;
see, e.g.,~\cite{LinP86,War97,Cha08,ChaSMQ10}.

Assembling the above expressions then yields the final evolution equation
\beq
\label{eq:sigevol2}
\frac{\partial \Sigma}{\partial t} = \frac{1}{r} \frac{\partial}{\partial r}
\left[
3 r^{1/2} \frac{\partial}{\partial r} \left( r^{1/2} \nu \Sigma \right)
- \frac{2 \Lambda \Sigma r^{3/2}}{M^{1/2}}
\right] \ .
\eeq

The accretion rate through any radius $r$ in the disk may be calculated from
\begin{eqnarray}
\label{eq:mdot}
\dot M (t,r) &=& 2 \pi r \Sigma (-v_r) \nonumber \\
       &=& - \left[ \frac {\partial (r^2 \Omega)}{\partial r} \right]^{-1} 
\frac{\partial G} {\partial r}\ .
\end{eqnarray}
Combining Eqs.~(\ref{eq:sigevol1}) and ~(\ref{eq:mdot}) yields
\begin{equation}
\label{eq:mdot2}
\frac{\partial \Sigma}{\partial t} 
=  \frac{1}{2 \pi r}
\frac{\partial}{\partial r} \dot M\ .
\end{equation}

To solve the above evolution equation we
impose the following boundary conditions:
\begin{equation}
\label{eq:bc}
\mbox{b.c.'s}: \ \ \ \nu \Sigma = \left \{ \begin{array}{ll}
(\nu \Sigma)_{\rm out},&
r = r_{\rm out}\\
 0,& r = r_{\rm isco}
\end{array} \right. .
\end{equation}
In Eq.~({\ref{eq:bc}) $r_{\rm out}$ is the outer radius
of the disk and $r_{\rm isco}$ is the ISCO radius of the primary.
Typically, $r_{\rm out} \gg r_{\rm isco}$ and in some cases one can take
$r_{\rm out} \rightarrow \infty$. We retain the solution for finite
$r_{\rm out}$ in part to facilitate numerical implementation of the outer
boundary condition.

\subsection{Orbital Evolution}

The rate at which the secondary black hole migrates inward is determined both
by back-reaction to the tidal torquing of the disk and by gravitational wave
emission~\cite{LodNKP09,ChaSMQ10}: 
\beq
\label{eq:aevol}
da/dt = (da/dt)_{\rm tid} + (da/dt)_{\rm GW}\ ,
\eeq
where
\beq
\label{eq:atidevol}
(da/dt)_{\rm tid} = -\frac{4 \pi a^{1/2}}{M^{3/2}q}
\int_{r_{\rm isco}}^{r_{\rm out}} r \Lambda \Sigma dr
\ \ \left( \equiv \frac{a}{t_{\rm tid}} \right)\ ,
\eeq
and where $(da/dt)_{\rm GW}$ is given by the familiar quadrupole-radiation 
orbital decay law, 
\begin{equation}
\label{eq:GW}
(da/dt)_{\rm GW} = -\frac{16}{5} \frac{M^3 \zeta}{a^3} \ \ \left( \equiv 
\frac{a}{t_{\rm GW}} \right)\ ,
\end{equation}
where $\zeta \equiv 4q/(1+q)^2$.
In Eq.~(\ref{eq:atidevol}) the integration is over the entire disk, although
most of the contribution from tidal torques
arises close to the gap boundaries near $ r \approx a$.

During any epoch in which back-reaction to tidal torques is not important, 
Eq.~(\ref{eq:aevol}) can be integrated to yield
\begin{equation}
\label{eq:anamerge}
a(t)/a(0) = \left( 1 - 4 t/t_{\rm GW}(0) \right)^{1/4}, 
\ \ t_{\rm GW}/t_{\rm tid} \ll 1\ ,
\end{equation}
where $t=0$ marks the beginning of such an epoch.

\subsection{Electromagnetic Radiation}

The local radiated emission from the disk arises both from viscous and tidal
dissipation. We assume that all of the dissipation is radiated locally, whereby
the local electromagnetic flux $F(t,r)$ from each side of the disk is equal
to the local dissipation rate $D(t,r)$ per unit surface area.
The rate of viscous dissipation is~\cite{Pri81}
\begin{equation}
\label{diss}
D_{\rm vis}(t,r)=\frac{9}{8}\nu \Sigma \frac{M}{r^3} = F_{\rm vis}(t,r).
\end{equation}
The rate of tidal dissipation is given by~\cite{GooR01, LodNKP09}}
\begin{equation}
\label{eq:Dtid}
D_{\rm tid}(t,r) = 
\frac{1}{2} \left( \Omega(a)-\Omega(r) \right) \Lambda \Sigma
= F_{\rm tid}(t,r).
\end{equation}

The local flux generates the luminosity $L(t,r)$ according to
\begin{eqnarray}
\label{eq:Flxvis}
\hat{F}_{\rm vis}(t,r)  &\equiv& M^2 F_{\rm vis}(t,r)/\dot{M}_{\rm eq} 
\nonumber \\
          &=&\frac{1}{4 \pi}\left( \frac{M}{r} \right)^2 
         \frac{d} {{d \ln r}} \left( L_{\rm vis}(t,r)/\dot{M}_{\rm eq} \right), 
\end{eqnarray} 
with a similar relation between $\hat{F}_{\rm tid}(t,r)$ and 
$L_{\rm tid}(t,r)$. The total local flux is then given by 
\begin{equation}
\label{eq:Flx} 
\hat{F}(t,r) = \hat{F}_{\rm vis}(t,r) + \hat{F}_{\rm tid}(t,r),
\end{equation}
and the total luminosity integrated over the entire disk (both sides) is
\begin{equation}
\label{eq:lum}
L(t) = L_{\rm vis}(t) + L_{\rm tid}(t). 
\end{equation}
In Eq.~(\ref{eq:Flxvis}) $\dot{M}_{\rm eq}$ is the accretion rate in
an {\it equilibrium} disk about a {\it single} black hole of mass $M$.
When we compare Newtonian and GR-hybrid results we will use
Eq.~(\ref{eq:mdot5}) for $\dot{M}_{\rm eq}$ in all of our normalizations.

\subsection{Quasistationary Solution: Pre-Decoupling} 
\label{sec:IID}

As the binary inspiral proceeds from large separation, the 
inspiral timescale due to gravitational wave emission eventually 
becomes shorter than the viscous timescale in the disk, at which time 
the binary decouples from the disk and ultimately merges. We define the
decoupling radius $a_d$ to be the separation at which the two 
timescales become equal.  Prior to BHBH-disk 
decoupling the balance between tidal and
viscous torques drives the disk to a quasistationary equilibrium state,
perturbed slightly by small amplitude, spiral density waves emanating
from the edges of the gap. Previously we solved the disk evolution equations
in steady state to determine the quasistationary, (orbit-averaged)  surface
density profile prior to decoupling as a function of the the binary 
separation~\cite{LiuS10}; see also~\cite{KocHL12,Raf12}.
For these early epochs we
set $a =$ constant and $\partial \Sigma / \partial t = 0$
in Eq.~(\ref{eq:sigevol2}) to obtain the density profile. This quasistationary 
solution is used below as initial data for the Newtonian time-dependent 
simulations that evolve the binary-disk system from pre- 
to post-decoupling, continuing all the way through 
the late inspiral, merger and post-merger phases.
  
The accretion rate $\dot M$ in steady state is independent of $r$.
In steady state Eq.~({\ref{eq:sigevol2}) admits a first integral which,
when combined with Eq.~(\ref{eq:mdot}), yields a first-order ODE,
\begin{equation}
\label{eq:mdot3}
\dot M = 2 \pi \left[ 
3 r^{1/2}\frac{d \left( r^{1/2} \nu \Sigma \right)} {d r} 
- \frac{2 \Lambda \Sigma r^{3/2}}{M^{1/2}} 
\right ] 
= {\rm constant}.
\end{equation}
We could solve the second-order elliptic equation obtained by setting
the right-hand side of Eq.~(\ref{eq:sigevol2}) to zero
to obtain the steady-state density profile, then evaluate
Eq.~(\ref{eq:mdot3}) for the accretion rate. Alternatively, 
we could integrate the first-order Eq.~(\ref{eq:mdot3}) directly for the
density, which, when the boundary conditions are implemented, 
automatically provides  $\dot M$ as an eigenvalue. 
We chose the later strategy in~\cite{LiuS10}, but adopt the
former approach in Section~\ref{sec:IVB1} in obtaining  
the initial data.  

\subsection{Quasistationary Solution: Post-Merger}

Following binary merger the tidal torque vanishes while gas in the disk 
continues to diffuse inward on
a viscous timescale, accreting onto the remnant black hole and ultimately settling into
a final, stationary equilibrium state.
This stationary disk configuration is described by  well-known
analytic density and temperature profiles, as well as analytic local 
fluxes and distant total luminosities, and these quantities provide 
useful checks on the late stages of any
disk evolution calculation.  The final equilibrium density profile, 
obtained by integrating Eq.~(\ref{eq:sigevol2}) 
in steady-state in the absence of the tidal torque, is given by the familiar
result for a Shakura-Sunyaev Newtonian thin disk around a single black hole 
(see, e.g.,~\cite{FraKR02} and references therein), generalized
for a disk of finite radial extent~\cite{LiuS10}:
\begin{eqnarray}
\label{eq:Sigeq}
\nu \Sigma(r) &=& \left( \nu \Sigma \right)_{\rm out} 
\frac{\left( 1- r_{\rm isco}^{1/2}/ r^{1/2} \right )}
{\left( 1- r_{\rm isco}^{1/2}/ r_{\rm out}^{1/2} \right )}, 
\ \ \  [{\rm post-merger}] \cr
          &=& \frac{\dot M_{\rm eq}}{3 \pi} 
\left( 1 - r_{\rm isco}^{1/2}/r^{1/2} \right). 
\end{eqnarray}
The second equality above thus yields the steady-state accretion rate
$\dot M_{\rm eq}$ in terms of the density and viscosity at the outer
boundary:
\begin{equation}
\label{eq:mdot6}
\dot M_{\rm eq} = 3 \pi \nu_{\rm out} \Sigma_{\rm out} 
\frac{1}{1-r_{\rm isco}^{1/2}/r_{\rm out}^{1/2}}. 
\end{equation}
The corresponding stationary flux, due entirely to viscous dissipation, may be expressed
as  
\begin{equation}
\label{eq:Feq}
\hat{F}_{\rm vis}(r)=\frac{3}{8 \pi} \left( \frac{M}{r} \right)^3 
           \left( 1-r_{\rm isco}^{1/2}/r^{1/2} \right).
\end{equation}
The flux, together with Eq.~(\ref{eq:Flxvis}), gives the differential luminosity, 
\begin{equation}
\label{eq:dLeq1}
 \frac{d} {{d \ln r}} \left( L_{\rm vis}(r)/\dot{M}_{\rm eq} \right) =
\frac{3}{2} \frac{M}{r} \left( 1-r_{\rm isco}^{1/2}/r^{1/2} \right).
\end{equation}
The total steady-state luminosity integrated over the entire disk is then given by
\begin{eqnarray}
\label{eq:Leq}
L_{\rm vis} &=& \frac{\dot M_{\rm eq} M}{2 r_{\rm isco}} 
\left ( 1 - 3 \frac{r_{\rm isco}}{r_{\rm out}}
+2 \frac{r_{\rm isco}^{3/2}}{r_{\rm out}^{3/2}} \right), 
\ \ \  [{\rm post-merger}] \cr
  &=& \frac{\dot M_{\rm eq} M}{2 r_{\rm isco}},  \ \ \ 
r_{\rm out} \rightarrow \infty.
\end{eqnarray}

\section{The GR-Hybrid Evolution Equations}
\subsection{Disk Evolution}
\label{sec:IIIA}

We propose the following GR-hybrid evolution equation for the rest-mass 
surface density  ($\Sigma \equiv \int \rho_0 dz$, where $\rho_0$ is the
rest-mass density) to replace Eq.~(\ref{eq:sigevol2}):
\beq
\label{eq:sigevol3}
\frac{\partial \Sigma}{\partial t} = \frac{1}{\Gamma r} 
\frac{\partial}{\partial r} \left[ \frac{\Gamma}{Q}
3 r^{1/2} \frac{\partial}{\partial r} 
\left( r^{1/2} \nu \Sigma \frac{\mathscr{D}^2}{\mathscr{C}} \right)
- \frac{2 \Lambda \Sigma r^{3/2}}{M^{1/2}}
\right],
\eeq
In assembling the above equation
we specialized to the Kerr metric in Boyer-Lindquist coordinates to
describe the (quasi-) stationary spacetime established by the 
more massive primary black hole. We used this metric 
to express the following functions, many of which were introduced by 
Novikov and Thorne~\cite{NovT73} (see also Page and Thorne~\cite{PagT74}):
\begin{eqnarray}
\label{eq:Kerr}
M &=& {\rm mass \  of \ primary \ black \ hole}, \cr
J &=& {\rm \ spin \ angular \ momentum \ of \ the \ primary \ hole}, \cr
a_* &=& J/M^2, \ \ \ 0 \leq a_* \leq 1, \ \ \cr
x &=& (r/M)^{1/2}, \cr
\Gamma &=& \mathscr{B} \mathscr{C}^{-1/2}, \cr 
L^+ &=& M x \mathscr{C}^{-1/2} (1 - 2 a_* x^{-3} + a_*^2 x^{-4}), \cr
Q &=& 2 x^{1/2} \partial L^+/\partial r, \cr
\mathscr{B} &=& 1 + a_* x^{-3}, \cr
\mathscr{C} &=& 1 - 3x^{-2} + 2 a_* x^{-3}, \cr
\mathscr{D} &=& 1 - 2 x^{-2} + a_*^2 x^{-4}, \cr
\mathscr{G} &=& 1 - 2 x^{-2} + a_* x^{-3}, \cr 
\mathscr{Q} &=& \ {\rm Eq}.~(35) \ \text{in \cite{PagT74}}, \cr
\mathscr{R} &=& \mathscr{Q}/\mathscr{B}.  
\end{eqnarray}

We note that in the case of a thin disk around a stationary Kerr black hole
$\Sigma$ as defined above is a scalar invariant, like $\rho_0$. 

Our proposed disk evolution Eq.~(\ref{eq:sigevol3}) has the following features:

\begin{enumerate}
\setcounter{enumi}{0}

\item We assume that during all epochs the gas flow takes place in the 
background geometry of the more massive primary, which we approximate
by the stationary Kerr metric.
The viscous torque, described by the first term on the right-hand side,
is treated in full GR for gas flow in a thin Keplerian disk. In the 
absence of the tidal torque term arising from the presence of the secondary,
the equation reduces identically to the evolution equation presented in 
Lightman and Eardley~\cite{LigE75} for time-dependent disk accretion
onto a single Kerr black hole. 

\item The tidal torque density, accounted for by the second term on the
right-hand side, is based on the Newtonian formula, Eq.(~\ref{eq:AN}).

\item The entire equation reduces identically to Eq.~(\ref{eq:sigevol2}) in the
weak-field region, i.e., for $r/M \gg 1$, whereby $\Gamma, 
\mathscr{D}, \mathscr{C}$ and $Q$ all appproach unity.

\item In the absence of tidal torques,  
the steady-state solution of Eq.~(\ref{eq:sigevol3}) gives 
the same profile for the product $\nu \Sigma$ that characterizes a 
standard relativistic Novikov-Thorne accretion disk around a single 
Kerr black hole [again generalized for a disk of finite radial extent; 
see Eq.~(\ref{eq:Sigeq2})].

\end{enumerate}

Basing the tidal torque on the Newtonian expression is motivated by the fact
that this torque is strongly peaked near the orbit of the secondary 
and plays its most important role when the binary separation
$a$ is large. Specifically, the disk radii in which the 
torque is most significant
typically satisfy $r \sim a \gg M$ and thus reside in the weak-field region
outside the primary.  (We neglect any accretion onto the secondary,
which is expected to be small). 
By contrast, the viscous torque drives gas into the strong-field region and 
into the primary during all epochs and this flow requires a full GR treatment 
Moreover, following merger,
the tidal term vanishes and Eq.~(\ref{eq:sigevol3}) 
reliably accounts for the inward diffusion of gas, 
the filling of any pre-merger gaps in the disk, the time-varying 
accretion onto the remnant, and the
relaxation of the disk and accretion rate to a (quasi-)stationary state, 
all in full GR.

A more rigorous treatment would incorporate a
relativistic tidal torque density $dT_{\rm tid}/dr$ in place of the
Newtonian expression used here.
Such a relativistic formula presumably can be obtained by
employing the relativistic tidal torque 
derived by Hirata~\cite{Hir11a,Hir11b} for a single Lindblad resonance
in an accretion disk that orbits a Kerr black hole and is perturbed
by a small secondary. This formula may be summed over
many resonances, treated as a continuum, to get a smooth torque density.
Such a sum has been carried out only for a Newtonian disk~\cite{GolT80},
and has been used here.
However, as described above, employing this Newtonian
formula in a first approximation  should be adequate to treat many of 
the epochs of interest during the merger event.

To solve Eq.~(\ref{eq:sigevol3}) we impose the same boundary conditions
as specified by Eq.~(\ref{eq:bc}). The radius $r_{\rm isco}$ is given
by the familiar expressions for a Kerr black hole in 
Boyer-Lindquist coordinates (see,e.g. \cite{ShaT83}, Eq.~12.7.24}). 

The {\it rest-mass} accretion rate through any radius $r$ is given by
\begin{eqnarray}
\label{eq:mdot4}
{\dot M}_0(t,r) &=& 2 \pi r \Sigma (-v^{\hat{r}}) \mathscr{D}^{1/2} \\
   &=& 2 \pi \left[
\frac{\Gamma}{Q}
3 r^{1/2} \frac{\partial}{\partial r}
\left( r^{1/2} \nu \Sigma \frac{\mathscr{D}^2}{\mathscr{C}} \right)
- \frac{2 \Lambda \Sigma r^{3/2}}{M^{1/2}} 
\right ] \nonumber . 
\end{eqnarray}
Combining Eqs.~(\ref{eq:sigevol3}) and (\ref{eq:mdot4}) then yields
\begin{equation}
\label{eq:sigevmdot}
\frac{\partial \Sigma}{\partial t} 
=  \frac{1}{2 \pi r \Gamma}
\frac{\partial}{\partial r} \dot M_0.
\end{equation}
Eqs.~(\ref{eq:mdot4}) and (\ref{eq:sigevmdot}) reduce to Eqs.~(\ref{eq:mdot})
and (\ref{eq:mdot2}) in the Newtonian limit.

\subsection{Orbital Evolution}
\label{sec:IIIB}

The inspiral of a low-mass black hole companion onto a more massive
primary is a nontrivial problem in general relativity, even in vacuum. 
Post-Newtonian (PN) approaches, based on expansions in $v^2/c^2$, can treat
most of the inspiral epochs, but break down once 
the orbital separation shrinks to within a few 
times the radius of the massive primary. 
Treating the problem without approximation using the tools of numerical
relativity is not computationally practical for following the 
inspiral from large separations characterizing binary-disk decoupling, 
but it can match onto PN 
trajectories at late times to continue the late-inspiral motion through
plunge, merger and ringdown (for an overview and references, see 
\cite{BauS10}). However, numerical relativity cannot yet 
evolve binaries with mass ratios $q < 10^{-2}$ because of the 
excessive dynamic range and resulting resolution requirements. 
For such small mass ratios black hole perturbation theory provides 
the best approximation, although it is computationally and analytically 
expensive. One approach involves the calculation of the self-field
acting on a test particle and following how it alters the orbital
trajectory (For a status report and references see~\cite{PoiPV11}).
A simpler, but more approximate, method is the ``radiative-adiabatic''
scheme, where the inspiral is treated as a sequence of adiabatically shrinking
geodesics. The shrinkage is determined by effectively calculating the
rate of change of the constants of the motion (energy, angular momentum and 
Carter constant) due to GW emission. (For a summary and references 
see~\cite{YunBHPBMT11}). Simplifications arise for the case of interest
here, where the orbit is nearly circular and resides in the equatorial
plane of the primary.

It is therefore possible to modify Eq.~(\ref{eq:GW}) to obtain a 
more reliable expression for the orbital decay due
to GW emission that accounts for higher-order general relativitistic
effects. Here, however, we
will continue to use Eq,~(\ref{eq:GW}) for simplicity, as it is adequate 
to illustrate disk evolution via the GR-hybrid approach in a first 
approximation and can be
generalized in subsequent analyses using the GR methods summarized above. 
Moreover, the deviations from the
simple quadrupole approximation that arise
when the secondary approaches the primary only lasts for a brief time 
interval, during which the bulk of the disk barely alters its structure.
For the same reasons we will continue to use
Eq.~(\ref{eq:atidevol}) to approximate the back-reaction of the tidal
torque on the companion.

\subsection{Electromagnetic Radiation}

The local radiation flux $F_{\rm vis}^{\rm com}(t,r)$  
removes the local viscous dissipation in the disk. Measured from each side 
of the disk per unit surface area by an 
observer comoving with the gas, it is given by
\begin{equation}
D_{\rm vis} = F_{\rm vis}^{\rm com}(t,r)=\frac{3}{4}\left( \frac{M}{r^3} 
\right)^{1/2} \mathscr{W}(t,r) \frac{\mathscr{D}}{\mathscr{C}}, 
\end{equation}
where $\mathscr{W}(t,r)$ is the vertically-integrated
shear stress~\cite{NovT73},
\begin{equation}
\mathscr{W}(t,r) = \int dz t_{\hat{\phi} \hat{r}} = 
\frac{3}{2}\nu \Sigma \frac{M^{1/2}}{r^{3/2}}
\frac{\mathscr{D}}{\mathscr{C}}.
\end{equation}
We then define
\begin{eqnarray}
\label{eq:fluxhat2}
\hat{F}_{\rm vis}(t,r)  &\equiv& M^2 F_{\rm vis}^{\rm com}(t,r)/\dot{M}_{\rm eq} \cr
&=& \frac{9}{8}\nu \Sigma \left( \frac{M}{r} \right)^3
\left( \frac{\mathscr{D}}{\mathscr{C}} \right)^2/ \dot{M_{\rm eq}}, 
\end{eqnarray} 
where $\dot{M}_{\rm eq}$ is the rest-mass accretion rate in an equilibrium disk
about a single black hole of mass $M$ (see Eq.~\ref{eq:mdot5}).
Consistent with our adopting Newtonian approximation for the tidal 
torque and tidal dissipation, we again use Eq.~(\ref{eq:Dtid}) for $F_{\rm tid}$
and define $\hat{F}_{\rm tid}(t,r)$ by analogy with $\hat{F}_{\rm vis}(t,r)$.

Following~\cite{KulPSSNSZMDM11} we define a flux $F$ 
in terms of the radiated 
energy as measured by a distant, stationary observer: 
$F(t,r) \equiv dE/(r dr d\phi dt) = -u_t F^{\rm com}(t,r)$. Here 
$E$ is the energy measured by the distant observer and $u^{\alpha}$ is the 
4-velocity of a fluid element in a circular equatorial geodesic orbit 
about the primary: 
$u_t = - {\bf \tilde{\omega}^{\hat 0}}  \cdot \partial / \partial t = -\mathscr{G}/\mathscr{C}^{1/2}$. Now
the spacetime metric, which is dominated by the 
primary, is stationary. Moreover the orbit-averaged disk and its associated 
electromagnetic emission evolve on a slow, secular timescale 
($\sim {\rm min}[t_{\rm vis}, t_{\rm GW}]$) during most phases.
In this limit 
the total luminosity measured by a distant observer
may be computed from $L(t) \equiv 2 dE/dt = 4 \pi \int F(t',r') r' dr'$ or 
$d (L/\dot{M}_{\rm eq})/ d (\ln r) = 4 \pi r^2 F(t',r)/\dot{M}_{\rm eq}$,
where $t'$ is the retarded time from the observer to the source (the 
``fast-light'' approximation). We omit 
the small correction for any emitted radiation captured by the black holes.

\subsection{Quasistationary Solution: Pre-Decoupling}
\label{sec:IIID}

The discussion in Section~\ref{sec:IID} again applies.
For binary separations
$a \gg a_d$, the disk profile relaxes approximately 
to a quasistationary profile
found by setting $\partial \Sigma/ \partial t = 0$ in 
Eq.~(\ref{eq:sigevol3}) and solving the resulting elliptic equation.
We do so below in Section~\ref{sec:IVB1} to determine 
the initial data for an evolution calculation.
In this limit, the steady-state accretion rate $\dot{M}_{0}$ given 
by Eq.~(\ref{eq:mdot4}) satisfies
\begin{equation}
\label{eq:eigen}
\dot {M}_0 = 
 \ {\rm constant}. 
\end{equation}

\subsection{Quasistationary Solution: Post-Merger}

Following black hole merger, 
a transient epoch ensues, wherein the gas diffuses inward toward the primary 
according to Eq.~(\ref{eq:sigevol3}) in the absence of tidal torques and
fills in any gaps that had been generated by 
the secondary.  The disk eventually settles into a 
steady-state, relativistic Novikov-Thorne thin disk about the remnant black
hole, whereby the surface density satisfies $\partial \Sigma/ \partial t = 0$,
yielding 
\begin{eqnarray}
\label{eq:Sigeq2}
\nu \Sigma(r) &=& \left( \nu \Sigma \right)_{\rm out} 
\frac{\mathscr{C}^{3/2}}{\mathscr{C}_{\rm out}^{3/2}}                 
\frac{\mathscr{D}_{\rm out}^2}{\mathscr{D}^2}                 
\frac{\mathscr{R}}{\mathscr{R}_{\rm out}},
\ \ \ \  [{\rm post-merger}] \cr
&=& \frac{\dot M_{\rm eq}}{3 \pi} 
\frac{\mathscr{C}^{3/2}}
{\mathscr{D}^2}                 
\mathscr{R}.
\end{eqnarray}
Note that while Eq.~(\ref{eq:Sigeq2}) reduces to the Newtonian result, 
Eq.~(\ref{eq:Sigeq}), as $r \rightarrow \infty$,
the equilibrium profiles differ to $\mathcal{O}(M/r)^{1/2}$, 
a significant difference for gas near the remnant black hole.
The second equality in Eq.~(\ref{eq:Sigeq2}) gives the steady-state, rest-mass 
accretion rate $\dot{M}_{\rm eq}$ in terms of the density and viscosity at
the outer boundary:
\begin{equation}
\label{eq:mdot5}
\dot M_{\rm eq} = 3 \pi \nu_{\rm out} \Sigma_{out} 
\frac{\mathscr{D}^{2}_{\rm out}}{\mathscr{C}^{3/2}_{\rm out} 
\mathscr{R}_{\rm out}}.
\end{equation}

The gas moves in a nearly circular geodesic orbit
with an angular velocity
\begin{equation}
\Omega = \frac{M^{1/2}}{r^{3/2}} \frac{1}{\mathscr{B}},
\end{equation}
and an inward radial drift
\begin{equation}
v^{\hat r} = - \frac{3}{2} \frac{\nu}{r} 
\left( \frac{\mathscr{D}}{\mathscr{C}} \right)^{3/2} \frac{1}{\mathscr{R}},
\end{equation}
as measured in the orthonormal orbiting frame. 
Combining Eqs.~(\ref{eq:fluxhat2}) 
and (\ref{eq:Sigeq2}) yields 
the comoving stationary flux, which is now due entirely to viscous dissipation:
\begin{equation}
\label{eq:Feq2}
\hat{F}_{\rm vis}(r)=\frac{3}{8 \pi} \left( \frac{M}{r} \right)^3 
           \frac{\mathscr{R}}{\mathscr{C}^{1/2}}.
\end{equation}
The flux results in a steady-state, differential luminosity,
\begin{equation}
\label{eq:dLeq}
 \frac{d} {{d \ln r}} \left( L_{\rm vis}(r)/\dot{M}_{\rm eq} \right) =
\frac{3}{2} \frac{M}{r} \frac{\mathscr{G} \mathscr{R}}{\mathscr{C}}.
\end{equation}
Integrating Eq.~(\ref{eq:dLeq}) over the entire disk yields the total 
observed luminosity $L_{\rm vis}$. For an infinite disk this integration
yields
\begin{equation}
\label{eq:lumtot1}
L_{\rm vis}/\dot{M}_{\rm eq} = 1 - \tilde{E}_{\rm isco}
\ \ (\equiv \eta), \ \ \ r_{\rm out} \to \infty,
\end{equation}
where $\tilde{E}_{\rm isco}$ is the binding energy per unit mass 
of a test particle in a circular geodesic orbit at $r_{\rm isco}$,
\begin{equation}
\tilde{E}_{\rm isco} = \frac{r_{\rm isco}^2 - 2 M r_{\rm isco} + 
a_* M \sqrt{M r_{\rm isco}}}{r_{\rm isco}(r_{\rm isco}^2 -
  3 M r_{\rm isco} + 2 a_* M \sqrt{M r_{\rm isco}})^{1/2}}. 
\end{equation}
For a large, but finite disk with $M \ll r_{\rm out} < \infty$ we have
\begin{eqnarray}
\label{eq:lumtot2}
L_{\rm vis}/\dot{M}_{\rm eq} &\approx& 1 - \tilde{E}_{\rm isco} 
-2 \int_{r_{\rm out}}^\infty D_{\rm vis} 2 \pi r\ dr \cr
&\approx& 1 - \tilde{E}_{\rm isco} -
\frac{3}{2} \frac{M}{r_{\rm out}}.
\end{eqnarray}
The right-hand side of Eq.~(\ref{eq:lumtot1}) yields the well-known
efficiency $\eta$ of stationary accretion from an infinite disk
onto a Kerr black hole: 5.72\% for $a_* = 0$
and 42.3\% for $a_* = 1$. The efficiency is less for a finite disk,
as indicated by Eq.~(\ref{eq:lumtot2}).

\section{Numerical Evolution}

\subsection{Nondimensionalization}

To solve Eq.(\ref{eq:sigevol3}) numerically it is convenient to introduce
the same nondimensional variables defined in~\cite{LiuS10}:

\begin{eqnarray}
\label{eq:nondim}
s &=& (r/r_{\rm out})^{1/2}, \ \ s_1 = (a/r_{\rm out})^{1/2}, 
\ \ s_2 = (r_{\rm isco}/r_{\rm out})^{1/2}, \cr 
\ \ \bar {\Sigma} &=& \Sigma/\Sigma_{\rm out}, 
\ \ \bar{\nu}=\nu/\nu_{\rm out}, \ \ y = s \bar{\Sigma}, \cr 
\ \ \ \bar {h} &=& h/r, \ \ \tau=t/2 t_{\rm vis}(r_{\rm out}). 
\end{eqnarray}
Here 
\begin{equation}
t_{\rm vis}(r) = \frac{2}{3} \frac{r^2}{\nu}
\end{equation}
is the characteristic viscous timescale at radius $r$ in the disk.

In terms of these variables, equation~(\ref{eq:sigevol3}) becomes
\begin{equation}
\label{eq:ndsigevol3}
\frac{\partial y}{\partial \tau} = \frac{1}{\Gamma s^2} 
\frac{\partial}{\partial s} 
\left [ 
\frac{\Gamma}{Q} \frac{\partial}{\partial s} 
\left( \bar \nu y \frac{\mathscr{D}^2}{\mathscr{C}} \right )   
- \frac{ g^*(s) y}{\left( {\rm max} (|s^2 - s_1^2|, s^2 \bar {h}) \right)^4} 
\right ] 
\end{equation}
where
\beq
\label{eq:g}
  g^*(s) = \left \{ \begin{array}{ll}
        g s_1^8      & s>s_1 \\
        -g s^8     & s<s_1
        \end{array} \right. \ ,
\eeq
and where
\beq
\label{eq:g*}
g = \frac{2}{3} f q^2 \left( \frac{\nu_{\rm out}}{M} \right)^{-1} 
    \left( \frac{r_{\rm out}}{M} \right)^{1/2}
\eeq
Eq.~(\ref{eq:ndsigevol3}) must be solved for $s\in [s_2,1]$ subject 
to the boundary conditions
\begin{equation}
\label{eq:bc2}
\mbox{b.c.'s}: \ \ \ y = \left \{ \begin{array}{ll}
1,&
s = 1\\
 0,& s = s_2
\end{array} \right. .
\end{equation}

We integrate Eq.~(\ref{eq:ndsigevol3}) numerically, implementing 
a second-order, finite-difference, Crank-Nicholson scheme.
Such an approach allows for arbitrarily large time-steps without the
restriction of a Courant condition to insure stability. We choose a
logarithmically increasing grid in radius to cover the large 
dynamic range in the disk with adequate spatial resolution everywhere.

\subsection{A Numerical Example}

To illustrate how a BHBH-disk system evolves when governed by the
GR-hybrid equation we track a typical BHBH-disk system by integrating
this equation in time.
Following our approach in~\cite{LiuS10},  we take the 
viscosity to have a power-law profile $\nu(r) \propto r^n$.
We then specify the system by first choosing the
parameters $q, a/M, r_{\rm out}/M \gg 1$ and $n$. We determine the
decoupling separation $a_d$ by the condition
$t_{\rm GW}(a_d) = \beta t_{\rm vis}(2a_d)$ (setting $\beta=0.1$),
which yields
\begin{equation}
a_d/M = \left[ \frac{128}{15 \cdot 2^n}\beta \zeta
\left( \frac{\nu_{\rm out}}{M} \right)^{-1} 
\left(\frac{r_{\rm out}}{M}\right)^n\right]^{1/(n+2)}.
\label{eq:ad2}
\end{equation}
We next fix $\bar{h}=h/r =0.1$, which essentially establishes the disk
thickness near $r=a$, where it most matters. 
To set the scale for the density and disk size 
in physical units we fix $\Sigma_{\rm out}$ and $r_{\rm out}$, 
which determine the disk mass $M_{\rm disk}$.
Finally, we set $\nu_{\rm out}$ by specifying the final accretion
rate onto the black hole remnant for an {\it infinite} disk,
$\dot{M}_{\rm rem} = 3 \pi \nu_{\rm out} \Sigma_{\rm out}$
(see Eqs.~\ref{eq:mdot6} or \ref{eq:mdot5} in the limit 
$r_{\rm out} \gg M$)
to be a fraction $\gamma$ of the Eddington value,
$\gamma = {\dot M}_{\rm rem}/{\dot M}_{\rm Edd}$. Here  
${\dot M}_{\rm Edd} \equiv L_{\rm Edd}/\eta = 4 \pi M m_p/(\eta \sigma_T)$, 
where $m_p$ is the proton mass, $\sigma_T$ is the Thomson cross-section 
and $\eta$ is the radiative efficiency [see Eq.~(\ref{eq:lumtot1})]. This
condition yields
\begin{equation}
\nu_{\rm out}/M = \frac{4}{3} \frac{\gamma}{\eta} 
\frac{m_p}{\sigma_T \Sigma_{\rm out}}.
\end{equation}

As in~\cite{LiuS10}, we assign the values  
$n=0.5$ and $q=5 \times 10^{-3}$ 
and set $f=0.01$, $r_{\rm out} = 10^3M$, $\gamma=0.1$,
and $\Sigma_{\rm out}=1.5\times 10^4 {\rm g\, cm}^{-2}$. 
Our choice of parameters gives
$M_{\rm disk}/M_\odot \sim 5 \times 10^3 M_8^2$ ($M_8 \equiv M/10^8 M_\odot$),
which is safely smaller than the mass of the
secondary BH for all $M \lesssim 2 \times 10^{11} M_\odot$.
These values for the asymptotic disk density and
disk mass are comparable to cases considered in~\cite{ArmN02,ChaSMQ10}.
We also set $J/M^2 = 0.5$, which gives $r_{\rm isco}=4.23M$. {\it The value of the primary mass $M$ 
scales out of the problem}  when
solved in dimensionless form according to Eq.~(\ref{eq:ndsigevol3}).

The adopted parameters give a decoupling radius $a_d/M = 18.1$ and 
a dimensionless tidal torque parameter $g = 0.0194$ [see~\cite{LiuS10}, 
Eq. (35), for definition]. 

A more sophisticated treatment could adopt a
one-zone approach as in a Shakura-Sunyaev or Novikov-Thorne disk. 
There one employs a local
radiation prescription that yields the local temperature and pressure
for an $\alpha$-disk or $\beta$-disk viscosity law, and uses these 
to derive the viscosity and $h/r$ profiles self-consistently 
(see, e.g.~\cite{KocHL12, Raf12}).
However, the simplified, but physically plausible, assignments chosen here
are sufficient for illustrating the implementation of the 
GR-hybrid equation (where we merely take it out for a ``test-drive'')
and we postpone a more detailed analysis for a future investigation.

\subsubsection{Initial Data}
\label{sec:IVB1}

We start the evolution when the binary separation is at $a/M = 5 a_d/M = 90.3 $.
We thus begin to track the inspiral before binary-disk decoupling, when for 
much of the disk the evolution is still quasistationary. We can therefore set the initial
density profile $\Sigma$ to the quasistationary profile
found by setting $\partial \Sigma/ \partial t =0$ in Eq.~(\ref{eq:sigevol3}) 
(i.e., $\partial y/ \partial \tau =0$ in Eq.~\ref{eq:ndsigevol3}) 
and solving the resulting elliptic equation, as 
discussed in Section~\ref{sec:IIID}.

At $t=0$ we have $\tilde{g} = 9.68$, where 
the quantity  $\tilde{g}$ is defined by Eq.~(55) of~\cite{LiuS10} and 
is related to $g$ according to
\begin{equation}
\tilde{g} = \frac{1}{2} g \left( \frac{a}{r_{\rm out}} \right)^{(1/2-n)}
\left( \frac{h(a)}{a} \right)^{-3}\ .
\end{equation} 
As shown in that reference, $\tilde{g}$ measures the ratio
of the tidal to viscous torque on the disk at $r \sim a$.  Whenever
$\tilde{g} \gtrsim a \ few$ prior to merger 
the dimensionless accretion rate satisfies 
$\dot{m} \ll 1$, i.e.,  the 
accretion rate onto to the primary is much reduced below the value it 
would have in the absence of tidal torques from the secondary. 
At $t=0$ we find $m = 1.03 \times 10^{-2}$, consistent with this
expectation. Note that $\tilde{g}$ decreases rapidly with increasing
${\bar h}$. Hence for thicker disks with ${\bar h} \gtrsim 0.1$, 
${\tilde g}$ would be smaller and there could be little or 
no suppression of accretion by the secondary (cf.~\cite{KocHL12}). 

At $t=0$ we find $t_{\rm GW}/t_{\rm tid} = 0.0588 M_8$. This ratio,
which measures the relative importance of tidal to 
gravitational radiation back-reaction forces on the secondary,  
decreases as the inspiral proceeds. Hence tidal back-reaction is not
important in this particular scenario for all $M < 10^9$. Accordingly, we use
Eq.~(\ref{eq:anamerge}) to track the 
orbital separation, consistent with the discussion in Section~\ref{sec:IIIB}.

To assist in the interpretation of the numerical results and figures
below, we list the following
conversion of nondimensional to physical units applicable to this scenario: 

\begin{eqnarray}
t({\rm yrs}) &=& 0.7649 \times 10^5  M_8 \tau, \cr
a_d({\rm au}) &=& 17.84 M_8, \cr
{\dot M}_{\rm eq} (M_\odot \ {\rm yr^{-1}}) &=& \gamma {\dot M}_{\rm Edd} 
\mathscr{P}_{\rm out} = 0.3028 M_8, \cr 
L_{\rm eq}({\rm erg \ s^{-1}}) &=& \gamma L_{\rm Edd} \mathscr{P}_{\rm out} 
= 1.409 \times 10^{45} M_8.
\end{eqnarray}

The subscript 'eq' refers to final quasistationary values associated with 
accretion onto the black hole remnant. 
The factor $\mathscr{P}_{\rm out} \equiv \mathscr{D}^2_{\rm out}/
(\mathscr{C}^{3/2}_{\rm out} \mathscr{R}_{\rm out})$ 
corrects for a relativistic disk which is {\it finite} and 
not infinite, as is the case here ({\it cf.} Eq.~\ref{eq:mdot5}).

\subsubsection{Evolution}

\begin{figure}
\includegraphics[width=9cm]{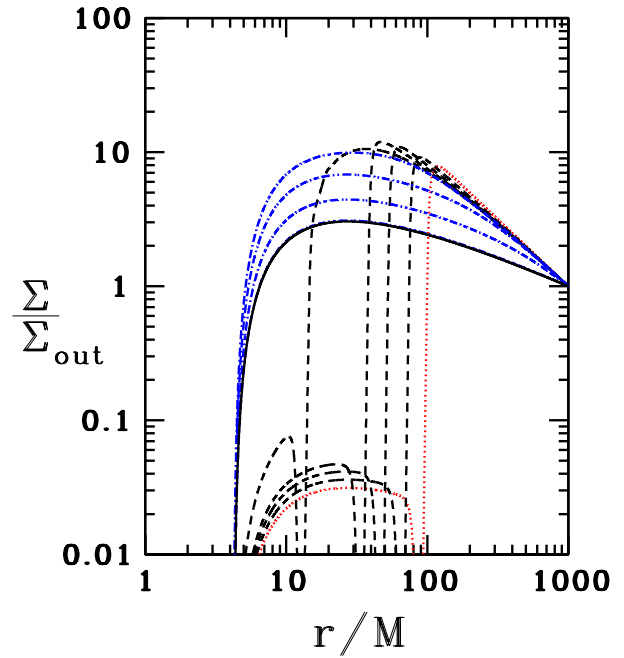}
\caption{
Snapshots of the disk surface density profile at selected times.
Profiles are shown for the initial disk at $\tau = 0$
({\it dotted red line}), the final equilibrium disk at
$\tau = \infty$ ({\it solid black line})
and for several intermediate times, $\tau$ =
0.03707, 0.04943, 0.05251, 0.05366  (pre-merger {\it dashed black lines}) and
0.05406, 0.06178, 0.09267, 0.2471 (post-merger {\it dot-dashed blue lines}).
Decoupling occurs at 
at $\tau$ = 0.05360 ($t = 4.099 \times 10^3 M_8$ yrs),
and merger occurs at $\tau$ = 0.05368 ($t = 4.106 \times 10^3 M_8$ yrs).
}
\label{fig:sigmaGR}
\end{figure}

The surface density profile is plotted at selected times during the evolution
in Fig.~\ref{fig:sigmaGR}. A gap forms in the disk near the orbital radius of
the secondary at $r = a(t)$ and moves with the secondary as it spirals inward. 
Disk-binary decoupling occurs after $a(t)$ has reached $a_d$, which happens at 
$\tau = 0.05360$ ($t=4.099 \times 10^3 M_8$ yrs). 
Prior to that time, the density
profile evolves in a quasistationary manner and remains close to the
quasistationary  solution
obtained by setting $\partial \Sigma / \partial t = 0$ at each 
orbital separation $a(t) > a_d$. Tidal torques, which are strongest 
near the orbital radius of the secondary,  cause a pile-up of the 
matter and create a density peak just outside $r = a(t)$.
As the orbital radius shrinks, the peak surface density moves inward 
while steadily increasing.  After merger, which occurs at
$\tau = 0.05368$ ($t = 4.106 \times 10^3 M_8$ yrs or 
$\Delta t = 6.57 M_8$ yrs after decoupling), 
the tidal torques vanish altogether while the 
residual viscous torques drive the inward diffusion of gas toward the 
remnant black hole. This inward diffusion reduces the peak value of the 
density and eliminates the gap inside $r = a_d$ altogether. 
By $\tau = 0.2471$  ($\Delta t = 1.479 \times 10^4 M_8$ yrs after merger), 
the density profile is seen in the figure to be nearly
indistinguishable from the equilibrium
Novikov-Thorne solution for a relativistic disk accreting 
onto a single black hole, Eq.~(\ref{eq:Sigeq2}). The density at
the ISCO vanishes at all times, as it is one of the boundary conditions.

\begin{figure}
\includegraphics[width=9cm]{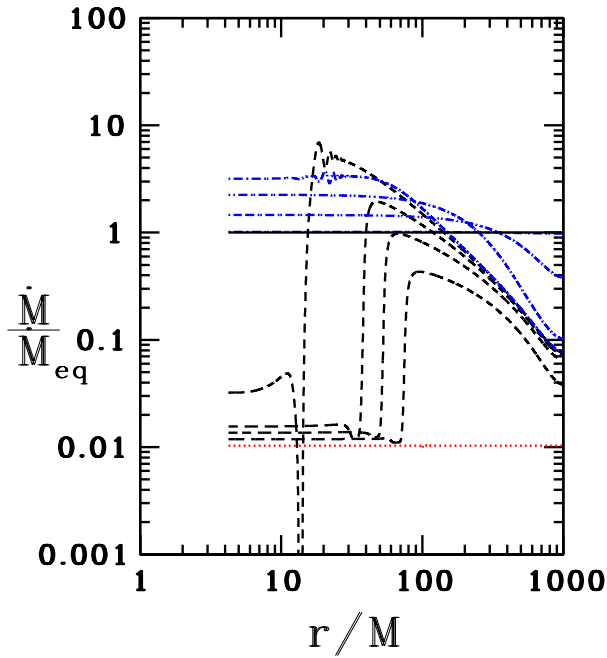}
\caption{
Snapshots of the rest-mass accretion rate at selected times.
Profiles are plotted at the same times shown in Fig.~\ref{fig:sigmaGR}.
}
\label{fig:mdotGR}
\end{figure}

The rest-mass accretion rate as a function of radius 
is plotted at select times in Fig.~\ref{fig:mdotGR}.
At $t=\tau=0$ the accretion rate, given by Eq.~(\ref{eq:mdot4}), is
everywhere constant. Such a result is a consequence of demanding that
$\partial \Sigma / \partial t = 0$ for the initial disk 
(see Eq.~\ref{eq:sigevmdot}). The normalized value of the initial
 accretion rate, $\dot M/\dot M_{\rm eq} = 1.031 \times 10^{-2}$, is much less
than unity, the value characterizing an equilbrium disk with the same
asymptotic density and viscosity about an isolated
black hole. As discussed above and in~\cite{LiuS10}, such a suppression 
of accretion flow is expected due to the high initial value of the torque
parameter  $\tilde g$.  As the inspiral proceeds, 
the exact cancellation of the viscous and tidal torque
terms in  Eq.~(\ref{eq:mdot4}) breaks down and the mass flux grows
behind the secondary. Soon after decoupling, the flow rate reaches values that 
actually {\it exceed} the final equilibrium value for an isolated 
black hole by almost a factor $\sim 10$ near $r \sim a_d$, 
as the tidal torques holding back the flow
diminish in strength and the density pile-up abates
(the ``bursting of the dam'').  Following merger the accretion 
rate eventually settles down to the constant 
equilibrium value, Eq.~(\ref{eq:mdot5}), throughout the disk.

\begin{figure}
\includegraphics[width=9cm]{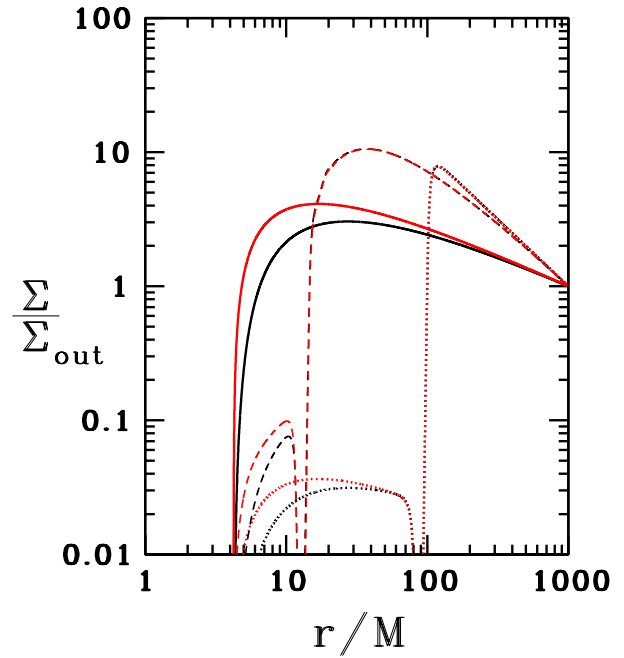}
\caption{
Comparison of GR-hybrid and Newtonian disk surface density profiles at selected
times. Profiles are shown for the initial disk at  $\tau = 0$
({\it dotted lines}), the final equilibrium disk at
$\tau = \infty$ ({\it solid lines}) and at one intermediate
time $\tau$ = 0.05366 (pre-merger dashed lines).
GR-hybrid lines are in {\it black}
and Newtonian lines in {\it red}; for 
each line type the lower (upper) curves are the
GR-hybrid (Newtonian) lines. 
Merger occurs at $\tau$ = 0.05368 ($t = 4.106 \times 10^3 M_8$ yrs).
}
\label{fig:sigmaGRNewt}
\end{figure}

It is interesting to compare the surface density profiles 
determined by integrating the
GR-hybrid and Newtonian evolution equations for the same
primary and secondary black hole masses,  disk 
parameters (i.e. $\Sigma_{\rm out}, \nu_{\rm out}, r_{\rm isco}, 
r_{\rm out}, n, f$ and $h/r$) and secondary orbit $a(t)$. 
In each case the initial data is
determined by setting $\partial \Sigma/\partial t = 0$ in their 
respective evolution equation and solving the resulting elliptic equation. 
The comparison is provided in Fig.~\ref{fig:sigmaGRNewt}. The initial 
quasistationary profiles are nearly identical, except in the strong-field
region $r/M \lesssim 20$, where the Newtonian profile is higher. 
(A comparison of profiles in the strong-field region is of course influenced by
gauge effects arising from the choice of radial coordinate, but $\Sigma$ is a
scalar invariant). The initial accretion rates are also slightly 
different (e.g. $\dot M/\dot M_{\rm eq} = 0.8892 \times 10^{-2}$ in the
Newtonian case), since
the elliptic equations that determine the rates are different in the
strong-field region.
Prior to merger but after decoupling, the density profiles outside the 
orbital radius are close, but inside that radius they continue to depart. 
After merger, the equilibrium profiles in the strong field regime remain 
different: the peak value of the surface density 
is higher in the Newtonian case by 35\%. The differences between the Newtonian
and GR-hybrid solutions become more pronounced as the primary spin increases 
and $a_* \rightarrow 1$.

\begin{figure}
\includegraphics[width=9cm]{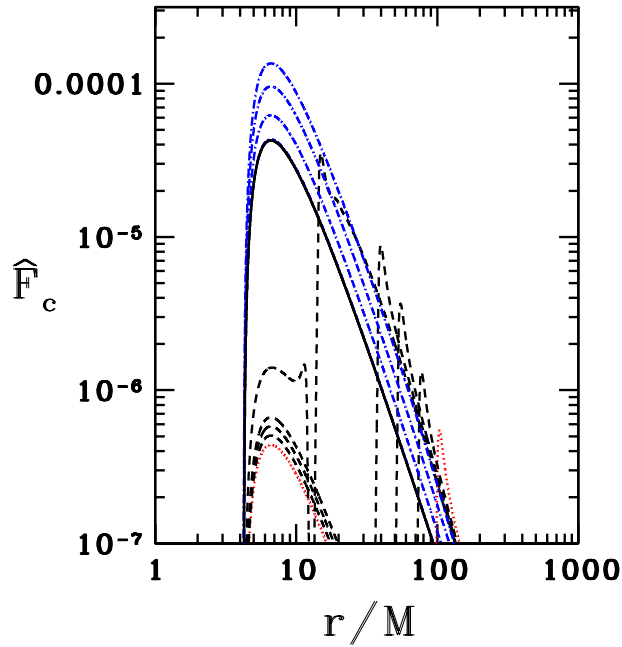}
\caption{
Snapshots of the comoving flux profile at selected times.
Profiles are plotted at the same times shown in Fig.~\ref{fig:sigmaGR}.
}
\label{fig:fluxGR}
\end{figure}

Profiles of the nondimensional comoving flux emerging from each
side of the disk are plotted at selected times in 
Fig.~\ref{fig:fluxGR}. The  evolution of the flux is correlated
with the evolution of the disk surface density. The peak flux occurs
just outside the orbital radius of the secondary prior to 
merger, and the regions where the flux dips correspond to the density
gaps near that orbit. Prior to merger, the tidal torque-driven density pile-up
causes the peak flux to increase with increasing time and decreasing
orbital radius, Following merger the tidal torque vanishes and 
the flux begins to decrease everywhere. 
Eventually the flux, now generated by viscous dissipation alone, 
settles into the equilibrium state corresponding to steady 
accretion of gas in a thin, relativistic disk onto a single black hole
[Eq.~(\ref{eq:Feq2})] The flux vanishes at the ISCO.

\begin{figure}
\includegraphics[width=9cm]{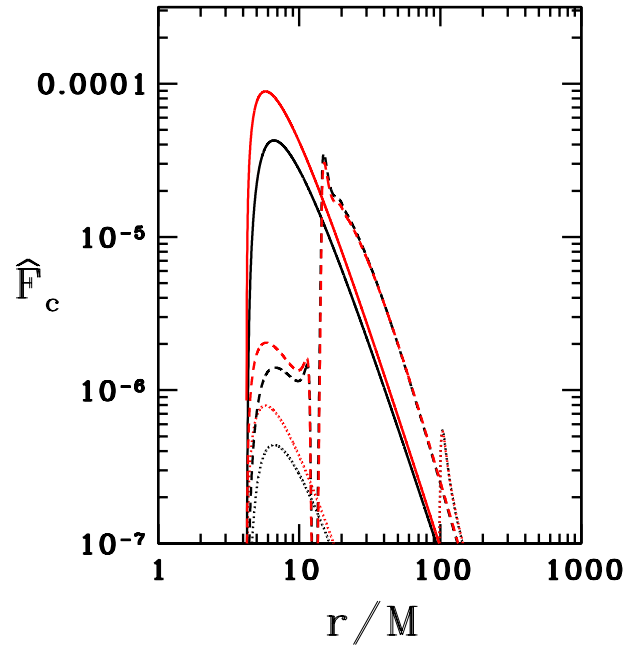}
\caption{
Comparison of GR-hybrid and Newtonian disk comoving flux profiles at selected
times. Profiles are plotted at the same times shown in
Fig.~\ref{fig:sigmaGRNewt}.
}
\label{fig:fluxGRNewt}
\end{figure}

A comparison of the comoving
fluxes determined from the GR-hybrid and Newtonian evolution equations
is presented in Fig.~\ref{fig:fluxGRNewt}. Prior to merger the flux profiles
are quite comparable but after merger the final equilibrium profiles
to which the disks relax differ significantly in the strong-field
region $r/M \lesssim 15$. The peak value of the final equilibrium comoving
flux is a factor of 2.1 times larger in the Newtonian case.

\begin{figure}
\includegraphics[width=9cm]{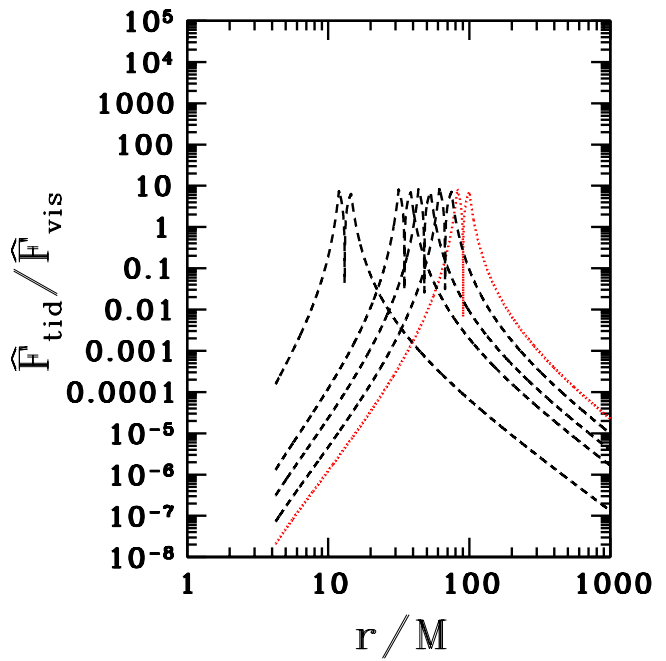}
\caption{
Snapshots of tidal-to-viscous comoving flux profiles at selected
times. Profiles are plotted at the same times shown in Fig.~\ref{fig:sigmaGR}
{\it prior} to merger; following merger tidal dissipation vanishes.
}
\label{fig:fluxratio}
\end{figure}

The ratio of the contribution of tidal heating to viscous heating 
to the comoving flux is plotted as a function of radius at selected
times prior to merger in Fig.~\ref{fig:fluxratio}. 
Tidal heating dominates over viscous heating  
near $r = a(t)$, but decreases 
rapidly both at smaller and larger
radii. [Note that exactly at $r = a(t)$ the tidal dissipation vanishes in 
accord with Eq.~(\ref{eq:Dtid}), hence the sudden dip in the curves].
This sudden fall-off is anticipated because the tidal torque, due to the
presence of the secondary, decreases rapdily with distance from the secondary. 

\begin{figure}
\includegraphics[width=9cm]{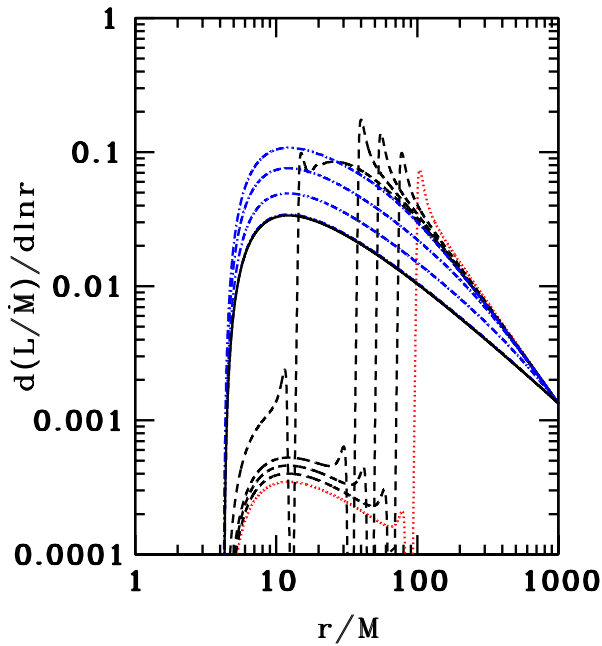}
\caption{
Snapshots of the differential luminosity profile, which shows the
contribution from various radii in the disk to the distant luminosity,
is plotted at selected times.  Profiles are plotted at the same times 
shown in Fig.~\ref{fig:sigmaGR}.
}
\label{fig:lum}
\end{figure}

The contribution from various radii in the disk 
to the luminosity measured by a distant observer is plotted in 
Fig.~\ref{fig:lum} at selected times. The evolution of this 
quantity follows the general trends already found
for the comoving flux shown in Fig.~\ref{fig:fluxGR}. The same differential
luminosity function (up to our normalization) has been
plotted in~\cite{KulPSSNSZMDM11} for 
relativistic, Novikov-Thorne thin disks undergoing steady-state 
accretion onto single Kerr black holes, where they are  
compared with three-dimensional GRMHD simulations of thin disks
($h/r \lesssim 0.1$) that relax to steady-state; see their Fig. 1.  
Our post-merger luminosities are all driven to these
Novikov-Thorne solutions at late times.
The GRMHD results are in general agreement with these solutions, 
but do exhibit some emission inside the ISCO, plus a small inward
shift of the peak emission to lower radii (see also~\cite{Sad09}).
These differences are not deemed significant enough to affect the accuracy
of, e.g.,  the continuum-fitting method, which employs the
Novikov-Thorne model to estimate black hole spins.

\begin{figure}
\includegraphics[width=9cm]{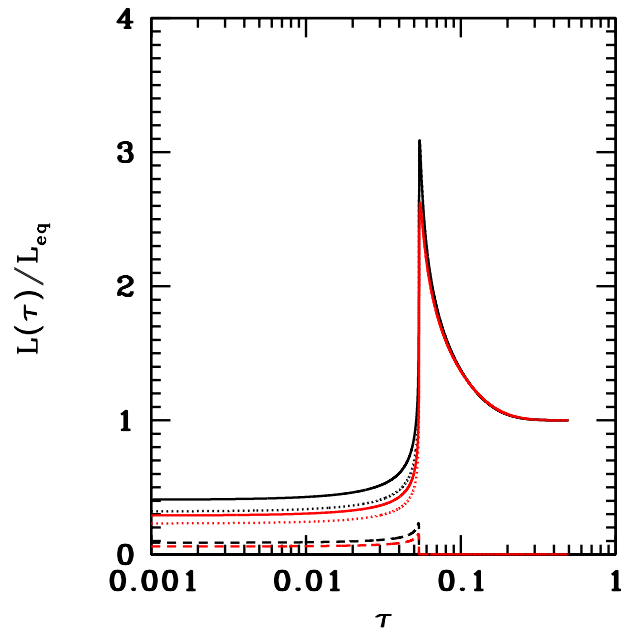}
\caption{
Variation of the distant 
total disk luminosity with time. The {\it dashed}
lines show the contribution from tidal dissipation, the {\it dotted}
lines from viscous dissipation. The {\it solid} lines show the total
luminosity. GR-hybrid lines are in {\it black}
and Newtonian lines in {\it red}; for each line type the upper (lower) 
curves are the GR-hybrid (Newtonian) lines. Merger occurs at $\tau$ = 0.05368
($t = 4.106 \times 10^3 M_8$ yrs).
}
\label{fig:lumtot}
\end{figure}

The evolution of the total electromagnetic 
luminosity from the disk measured by
a distant observer is plotted in Fig.~\ref{fig:lumtot}. There it
is seen that the tidal dissipation is always less important overall than
viscous dissipation and vanishes altogether following merger.
The luminosity rises sharply after merger, reaching a peak
at $\tau = 0.0542$ and decaying slowly thereafter to its final, equilibrium
value. The peak value is a full 3.08 times larger than its final
equilibrium value given by Eq.~\ref{eq:lumtot2},
$L_{\rm eq} = 0.0806 {\dot M}_{\rm eq}$. The rapid decline of the tidal torques
following decoupling and the sudden inward drift of matter from $r \sim a_d$ 
is responsible for the overshoot in luminosity. 
The full width at half-maximum of the luminosity curve 
is $\Delta \tau = 0.03126$ ($\Delta t = 2.39 \times 10^3 M_8$ yrs).
The rise, fall and asymptotic flattening of
the luminosity curve may provide an electromagnetic 
signature that a BHBH merger has occured in a circumbinary disk.

Also shown in  Fig.~\ref{fig:lumtot} are
the Newtonian evolution curves for the same quantities. The results
are qualitatively similar, but the equilibrium
accretion rates and radiation efficiencies are different from the
GR-hybrid solution (e.g., $L_{\rm eq} = 0.117 {\dot M}_{\rm eq}$, or 45\%
higher, in the
Newtonian case) and the peak luminosity overshoot is somewhat smaller 
(about 15\% lower in the Newtonian case).

\begin{figure}
\includegraphics[width=9cm]{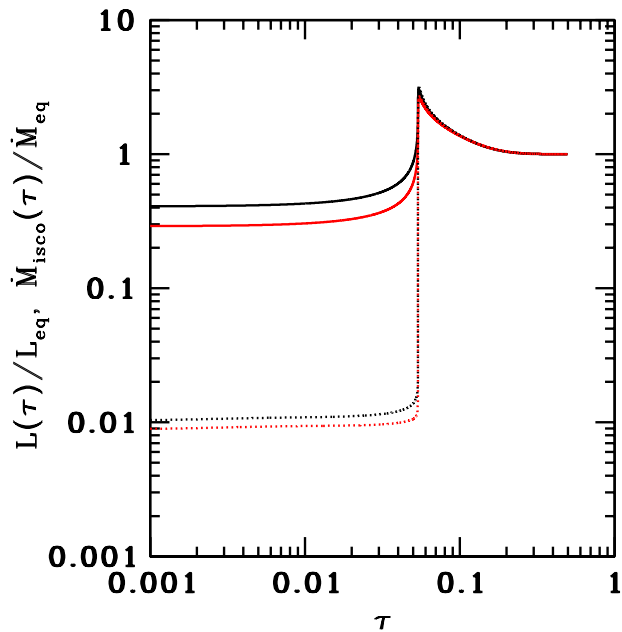}
\caption{
Variation of the distant total disk luminosity ({\it solid lines}) 
and ISCO accretion rate ({\it dotted lines}) with time. 
GR-hybrid lines are in {\it black}
and Newtonian lines in {\it red}; for each line type the upper (lower)
curves are the GR-hybrid (Newtonian) lines. Merger occurs at $\tau$ = 0.05368
($t = 4.106 \times 10^3 M_8$ yrs).
}
\label{fig:mdotlum}
\end{figure}

Fig.~\ref{fig:mdotlum} shows that the evolution of the total electromagnetic
luminosity is correlated with the evolution of the accretion rate at
the ISCO of the primary black hole. The overshoot of the accretion rate
above the final equilibrium value, and its subsequent decay to the
Novikov-Thorne value, drives the same time variation seen for 
the total luminosity.

As in the case of thin-disk accretion onto a single, stationary black hole,
Newtonian and GR models for low-mass BHBH-disk systems 
typically give the same {\it qualitative} results
for the observable EM radiation. Newtonian calculations are thus sufficient 
to identify characteristic EM luminosities and wavelengths. 
But the numerous factors of a $\sim$ few that comprise the {\it quantitative} 
differences between the models
are important if one hopes to use detailed observations to infer 
BH spins and other system parameters. For more complicated scenarios
than the one analyzed here there can even be  
qualitative differences involving relativistic effects 
that must be taken into account (e.g., binary remnant recoil, 
misaligned BH spins, accretion jets, etc.).

\section{Future Work}

The numerical scenario summarized here is presented as a simple demonstration
of the use of the GR-hybrid approach to track the orbit-averaged
evolution of a thin, Keplerian disk orbiting a low-mass BHBH, accounting
for some of the most important effects of general relativity. More
detailed microphysics, combined with a parameter survey, will be necessary 
to fully explore the consequences of this model. Future applications
should incorporate the following:

1. A self-consistent treatment of the viscosity and $h/r$ profiles by
implementing a Shakura-Sunyaev-Novikov-Thorne one-zone description of
each ring in the disk,
employing a local radiation prescription together with
an $\alpha$-disk or $\beta$-disk law for the viscosity to obtain the
required profiles;

2. A calculation of the observed radiation spectrum,  
by employing a ray-tracing or Monte-Carlo technique,
adapted to a time-dependent relativistic disk in curved spacetime 
(see, e.g.~\cite{DolGML09, SchH11, VinPGP11}). 

Several GR refinements can be implemented to improve the model while
retaining the spirit of an orbit-averaged description of an evolving 
BHBH-thin disk system. They include the following:

1. The replacement of the Newtonian formula with a
fully relativistic expression for the tidal torque density, 
along the lines discussed in Section~\ref{sec:IIIA};

2. Employing the true relativistic inspiral trajectory for
the low-mass BH companion, calculated using one of the GR techniques
discussed in Section~\ref{sec:IIIB}.

We intend to implement some of these improvements and apply the resulting
formalism in future studies.

{\it Acknowledgments}: It is a pleasure to thank C. Gammie and
V. Paschalidis for useful discussions and helpful comments.
This paper was supported in part by NSF Grant PHY09-63136 and 
NASA Grants NNX11AE11G and NNX13AH44G to the University of Illinois
 at Urbana-Champaign.

\bibliography{paper}

\begin{thebibliography}{10}%
\makeatletter
\providecommand \@ifxundefined [1]{%
 \ifx #1\undefined \expandafter \@firstoftwo
 \else \expandafter \@secondoftwo
\fi
}%
\providecommand \@ifnum [1]{%
 \ifnum #1\expandafter \@firstoftwo
 \else \expandafter \@secondoftwo
\fi
}%
\providecommand \enquote [1]{``#1''}%
\providecommand \bibnamefont  [1]{#1}%
\providecommand \bibfnamefont [1]{#1}%
\providecommand \citenamefont [1]{#1}%
\providecommand\href[0]{\@sanitize\@href}%
\providecommand\@href[1]{\endgroup\@@startlink{#1}\endgroup\@@href}%
\providecommand\@@href[1]{#1\@@endlink}%
\providecommand \@sanitize [0]{\begingroup\catcode`\&12\catcode`\#12\relax}%
\@ifxundefined \pdfoutput {\@firstoftwo}{%
 \@ifnum{\z@=\pdfoutput}{\@firstoftwo}{\@secondoftwo}%
}{%
 \providecommand\@@startlink[1]{\leavevmode\special{html:<a href="#1">}}%
 \providecommand\@@endlink[0]{\special{html:</a>}}%
}{%
 \providecommand\@@startlink[1]{%
  \leavevmode
  \pdfstartlink
   attr{/Border[0 0 1 ]/H/I/C[0 1 1]}%
   user{/Subtype/Link/A<</Type/Action/S/URI/URI(#1)>>}%
  \relax
 }%
 \providecommand\@@endlink[0]{\pdfendlink}%
}%
\providecommand \url  [0]{\begingroup\@sanitize \@url }%
\providecommand \@url [1]{\endgroup\@href {#1}{\urlprefix}}%
\providecommand \urlprefix [0]{URL }%
\providecommand \Eprint[0]{\href }%
\@ifxundefined \urlstyle {%
  \providecommand \doi [1]{doi:\discretionary{}{}{}#1}%
}{%
  \providecommand \doi [0]{doi:\discretionary{}{}{}\begingroup
  \urlstyle{rm}\Url }%
}%
\providecommand \doibase [0]{http://dx.doi.org/}%
\providecommand \Doi[1]{\href{\doibase#1}}%
\providecommand \bibAnnote [3]{%
  \BibitemShut{#1}%
  \begin{quotation}\noindent
    \textsc{Key:}\ #2\\\textsc{Annotation:}\ #3%
  \end{quotation}%
}%
\providecommand \bibAnnoteFile [2]{%
  \IfFileExists{#2}{\bibAnnote {#1} {#2} {\input{#2}}}{}%
}%
\providecommand \typeout [0]{\immediate \write \m@ne }%
\providecommand \selectlanguage [0]{\@gobble}%
\providecommand \bibinfo [0]{\@secondoftwo}%
\providecommand \bibfield [0]{\@secondoftwo}%
\providecommand \translation [1]{[#1]}%
\providecommand \BibitemOpen[0]{}%
\providecommand \bibitemStop [0]{}%
\providecommand \bibitemNoStop [0]{.\EOS\space}%
\providecommand \EOS [0]{\spacefactor3000\relax}%
\providecommand \BibitemShut [1]{\csname bibitem#1\endcsname}%
\bibitem{ArmN02}%
  \BibitemOpen
  \bibfield{author}{%
  \bibinfo {author} {\bibfnamefont{P.~J.}\ \bibnamefont{{Armitage}}}\ and\
  \bibinfo {author} {\bibfnamefont{P.}~\bibnamefont{{Natarajan}}},\ }%
  \bibfield{journal}{%
  \Doi{10.1086/339770}{\bibinfo {journal} {\apjl}}\ }%
  \textbf{\bibinfo {volume} {567}},\ \bibinfo {pages} {L9} (\bibinfo {month}
  {Mar.}\ \bibinfo {year} {2002})%
  \bibAnnoteFile{NoStop}{ArmN02}%
\bibitem{ChaSMQ10}%
  \BibitemOpen
  \bibfield{author}{%
  \bibinfo {author} {\bibfnamefont{P.}~\bibnamefont{{Chang}}}, \bibinfo
  {author} {\bibfnamefont{L.~E.}\ \bibnamefont{{Strubbe}}}, \bibinfo {author}
  {\bibfnamefont{K.}~\bibnamefont{{Menou}}},\ and\ \bibinfo {author}
  {\bibfnamefont{E.}~\bibnamefont{{Quataert}}},\ }%
  \bibfield{journal}{%
  \Doi{10.1111/j.1365-2966.2010.17056.x}{\bibinfo {journal} {\mnras}}\ }%
  \textbf{\bibinfo {volume} {407}},\ \bibinfo {pages} {2007} (\bibinfo {month}
  {Sep.}\ \bibinfo {year} {2010})%
  \bibAnnoteFile{NoStop}{ChaSMQ10}%
\bibitem{MilP05}%
  \BibitemOpen
  \bibfield{author}{%
  \bibinfo {author} {\bibfnamefont{M.}~\bibnamefont{{Milosavljevi{\'c}}}}\ and\
  \bibinfo {author} {\bibfnamefont{E.~S.}\ \bibnamefont{{Phinney}}},\ }%
  \bibfield{journal}{%
  \Doi{10.1086/429618}{\bibinfo {journal} {\apjl}}\ }%
  \textbf{\bibinfo {volume} {622}},\ \bibinfo {pages} {L93} (\bibinfo {month}
  {Apr.}\ \bibinfo {year} {2005})%
  \bibAnnoteFile{NoStop}{MilP05}%
\bibitem{RosLAPK09}%
  \BibitemOpen
  \bibfield{author}{%
  \bibinfo {author} {\bibfnamefont{E.~M.}\ \bibnamefont{{Rossi}}}, \bibinfo
  {author} {\bibfnamefont{G.}~\bibnamefont{{Lodato}}}, \bibinfo {author}
  {\bibfnamefont{P.~J.}\ \bibnamefont{{Armitage}}}, \bibinfo {author}
  {\bibfnamefont{J.~E.}\ \bibnamefont{{Pringle}}},\ and\ \bibinfo {author}
  {\bibfnamefont{A.~R.}\ \bibnamefont{{King}}},\ }%
  \bibfield{journal}{%
  \Doi{10.1111/j.1365-2966.2009.15802.x}{\bibinfo {journal} {\mnras}}\ }%
  \textbf{\bibinfo {volume} {401}},\ \bibinfo {pages} {2021} (\bibinfo {month}
  {Jan.}\ \bibinfo {year} {2010}),\
  \Eprint{http://arxiv.org/abs/0910.0002}{arXiv:0910.0002 [astro-ph.HE]}%
  \bibAnnoteFile{NoStop}{RosLAPK09}%
\bibitem{SchK08}%
  \BibitemOpen
  \bibfield{author}{%
  \bibinfo {author} {\bibfnamefont{J.~D.}\ \bibnamefont{{Schnittman}}}\ and\
  \bibinfo {author} {\bibfnamefont{J.~H.}\ \bibnamefont{{Krolik}}},\ }%
  \bibfield{journal}{%
  \Doi{10.1086/590363}{\bibinfo {journal} {\apj}}\ }%
  \textbf{\bibinfo {volume} {684}},\ \bibinfo {pages} {835} (\bibinfo {month}
  {Sep.}\ \bibinfo {year} {2008})%
  \bibAnnoteFile{NoStop}{SchK08}%
\bibitem{CorHM09}%
  \BibitemOpen
  \bibfield{author}{%
  \bibinfo {author} {\bibfnamefont{L.~R.}\ \bibnamefont{{Corrales}}}, \bibinfo
  {author} {\bibfnamefont{Z.}~\bibnamefont{{Haiman}}},\ and\ \bibinfo {author}
  {\bibfnamefont{A.}~\bibnamefont{{MacFadyen}}},\ }%
  \bibfield{journal}{%
  \Doi{10.1111/j.1365-2966.2010.16324.x}{\bibinfo {journal} {\mnras}}\ }%
  \textbf{\bibinfo {volume} {404}},\ \bibinfo {pages} {947} (\bibinfo {month}
  {May}\ \bibinfo {year} {2010}),\
  \Eprint{http://arxiv.org/abs/0910.0014}{arXiv:0910.0014 [astro-ph.HE]}%
  \bibAnnoteFile{NoStop}{CorHM09}%
\bibitem{OneMBRS09}%
  \BibitemOpen
  \bibfield{author}{%
  \bibinfo {author} {\bibfnamefont{S.~M.}\ \bibnamefont{{O'Neill}}}, \bibinfo
  {author} {\bibfnamefont{M.~C.}\ \bibnamefont{{Miller}}}, \bibinfo {author}
  {\bibfnamefont{T.}~\bibnamefont{{Bogdanovi{\'c}}}}, \bibinfo {author}
  {\bibfnamefont{C.~S.}\ \bibnamefont{{Reynolds}}},\ and\ \bibinfo {author}
  {\bibfnamefont{J.~D.}\ \bibnamefont{{Schnittman}}},\ }%
  \bibfield{journal}{%
  \Doi{10.1088/0004-637X/700/1/859}{\bibinfo {journal} {\apj}}\ }%
  \textbf{\bibinfo {volume} {700}},\ \bibinfo {pages} {859} (\bibinfo {month}
  {Jul.}\ \bibinfo {year} {2009})%
  \bibAnnoteFile{NoStop}{OneMBRS09}%
\bibitem{Sha10}%
  \BibitemOpen
  \bibfield{author}{%
  \bibinfo {author} {\bibfnamefont{S.~L.}\ \bibnamefont{{Shapiro}}},\ }%
  \bibfield{journal}{%
  \Doi{10.1103/PhysRevD.81.024019}{\bibinfo {journal} {\prd}}\ }%
  \textbf{\bibinfo {volume} {81}},\ \bibinfo {pages} {024019} (\bibinfo {month}
  {Jan.}\ \bibinfo {year} {2010}),\
  \Eprint{http://arxiv.org/abs/0912.2345}{0912.2345 [astro-ph.HE]}%
  \bibAnnoteFile{NoStop}{Sha10}%
\bibitem{TanM10}%
  \BibitemOpen
  \bibfield{author}{%
  \bibinfo {author} {\bibfnamefont{T.}~\bibnamefont{{Tanaka}}}\ and\ \bibinfo
  {author} {\bibfnamefont{K.}~\bibnamefont{{Menou}}},\ }%
  \bibfield{journal}{%
  \Doi{10.1088/0004-637X/714/1/404}{\bibinfo {journal} {\apj}}\ }%
  \textbf{\bibinfo {volume} {714}},\ \bibinfo {pages} {404} (\bibinfo {month}
  {May}\ \bibinfo {year} {2010}),\
  \Eprint{http://arxiv.org/abs/0912.2054}{arXiv:0912.2054 [astro-ph.CO]}%
  \bibAnnoteFile{NoStop}{TanM10}%
\bibitem{LodNKP09}%
  \BibitemOpen
  \bibfield{author}{%
  \bibinfo {author} {\bibfnamefont{G.}~\bibnamefont{{Lodato}}}, \bibinfo
  {author} {\bibfnamefont{S.}~\bibnamefont{{Nayakshin}}}, \bibinfo {author}
  {\bibfnamefont{A.~R.}\ \bibnamefont{{King}}},\ and\ \bibinfo {author}
  {\bibfnamefont{J.~E.}\ \bibnamefont{{Pringle}}},\ }%
  \bibfield{journal}{%
  \Doi{10.1111/j.1365-2966.2009.15179.x}{\bibinfo {journal} {\mnras}}\ }%
  \textbf{\bibinfo {volume} {398}},\ \bibinfo {pages} {1392} (\bibinfo {month}
  {Sep.}\ \bibinfo {year} {2009}),\
  \Eprint{http://arxiv.org/abs/0906.0737}{arXiv:0906.0737 [astro-ph.CO]}%
  \bibAnnoteFile{NoStop}{LodNKP09}%
\bibitem{LiuS10}%
  \BibitemOpen
  \bibfield{author}{%
  \bibinfo {author} {\bibfnamefont{Y.~T.}\ \bibnamefont{{Liu}}}\ and\ \bibinfo
  {author} {\bibfnamefont{S.~L.}\ \bibnamefont{{Shapiro}}},\ }%
  \bibfield{journal}{%
  \Doi{10.1103/PhysRevD.82.123011}{\bibinfo {journal} {\prd}}\ }%
  \textbf{\bibinfo {volume} {82}},\ \bibinfo {eid} {123011} (\bibinfo {month}
  {Dec.}\ \bibinfo {year} {2010}),\
  \Eprint{http://arxiv.org/abs/1011.0002}{arXiv:1011.0002 [astro-ph.HE]}%
  \bibAnnoteFile{NoStop}{LiuS10}%
\bibitem{KocHL12}%
  \BibitemOpen
  \bibfield{author}{%
  \bibinfo {author} {\bibfnamefont{B.}~\bibnamefont{{Kocsis}}}, \bibinfo
  {author} {\bibfnamefont{Z.}~\bibnamefont{{Haiman}}},\ and\ \bibinfo {author}
  {\bibfnamefont{A.}~\bibnamefont{{Loeb}}},\ }%
  \bibfield{journal}{%
  \Doi{10.1111/j.1365-2966.2012.22118.x}{\bibinfo {journal} {\mnras}}\ }%
  \textbf{\bibinfo {volume} {427}},\ \bibinfo {pages} {2680} (\bibinfo {month}
  {Dec.}\ \bibinfo {year} {2012}),\
  \Eprint{http://arxiv.org/abs/1205.5268}{arXiv:1205.5268 [astro-ph.HE]}%
  \bibAnnoteFile{NoStop}{KocHL12}%
\bibitem{ShaS73}%
  \BibitemOpen
  \bibfield{author}{%
  \bibinfo {author} {\bibfnamefont{N.~I.}\ \bibnamefont{{Shakura}}}\ and\
  \bibinfo {author} {\bibfnamefont{R.~A.}\ \bibnamefont{{Sunyaev}}},\ }%
  \bibfield{journal}{%
  \bibinfo {journal} {A\&A}\ }%
  \textbf{\bibinfo {volume} {24}},\ \bibinfo {pages} {337} (\bibinfo {year}
  {1973})%
  \bibAnnoteFile{NoStop}{ShaS73}%
\bibitem{Raf12}%
  \BibitemOpen
  \bibfield{author}{%
  \bibinfo {author} {\bibfnamefont{R.~R.}\ \bibnamefont{{Rafikov}}},\ }%
  \bibfield{journal}{%
  \bibinfo {journal} {ArXiv e-prints}}%
   (\bibinfo {month} {May}\ \bibinfo {year} {2012}),\
  \Eprint{http://arxiv.org/abs/1205.5017}{arXiv:1205.5017 [astro-ph.GA]}%
  \bibAnnoteFile{NoStop}{Raf12}%
\bibitem{MacM08}%
  \BibitemOpen
  \bibfield{author}{%
  \bibinfo {author} {\bibfnamefont{A.~I.}\ \bibnamefont{{MacFadyen}}}\ and\
  \bibinfo {author} {\bibfnamefont{M.}~\bibnamefont{{Milosavljevi{\'c}}}},\ }%
  \bibfield{journal}{%
  \Doi{10.1086/523869}{\bibinfo {journal} {\apj}}\ }%
  \textbf{\bibinfo {volume} {672}},\ \bibinfo {pages} {83} (\bibinfo {month}
  {Jan.}\ \bibinfo {year} {2008})%
  \bibAnnoteFile{NoStop}{MacM08}%
\bibitem{ShiKLH12}%
  \BibitemOpen
  \bibfield{author}{%
  \bibinfo {author} {\bibfnamefont{J.-M.}\ \bibnamefont{{Shi}}}, \bibinfo
  {author} {\bibfnamefont{J.~H.}\ \bibnamefont{{Krolik}}}, \bibinfo {author}
  {\bibfnamefont{S.~H.}\ \bibnamefont{{Lubow}}},\ and\ \bibinfo {author}
  {\bibfnamefont{J.~F.}\ \bibnamefont{{Hawley}}},\ }%
  \bibfield{journal}{%
  \Doi{10.1088/0004-637X/749/2/118}{\bibinfo {journal} {\apj}}\ }%
  \textbf{\bibinfo {volume} {749}},\ \bibinfo {eid} {118} (\bibinfo {month}
  {Apr.}\ \bibinfo {year} {2012}),\
  \Eprint{http://arxiv.org/abs/1110.4866}{arXiv:1110.4866 [astro-ph.HE]}%
  \bibAnnoteFile{NoStop}{ShiKLH12}%
\bibitem{DorHM12}%
  \BibitemOpen
  \bibfield{author}{%
  \bibinfo {author} {\bibfnamefont{D.~J.}\ \bibnamefont{{D'Orazio}}}, \bibinfo
  {author} {\bibfnamefont{Z.}~\bibnamefont{{Haiman}}},\ and\ \bibinfo {author}
  {\bibfnamefont{A.}~\bibnamefont{{MacFadyen}}},\ }%
  \bibfield{journal}{%
  \bibinfo {journal} {ArXiv e-prints}}%
   (\bibinfo {month} {Oct.}\ \bibinfo {year} {2012}),\
  \Eprint{http://arxiv.org/abs/1210.0536}{arXiv:1210.0536 [astro-ph.GA]}%
  \bibAnnoteFile{NoStop}{DorHM12}%
\bibitem{MegAFHLLMN09}%
  \BibitemOpen
  \bibfield{author}{%
  \bibinfo {author} {\bibfnamefont{M.}~\bibnamefont{{Megevand}}}, \bibinfo
  {author} {\bibfnamefont{M.}~\bibnamefont{{Anderson}}}, \bibinfo {author}
  {\bibfnamefont{J.}~\bibnamefont{{Frank}}}, \bibinfo {author}
  {\bibfnamefont{E.~W.}\ \bibnamefont{{Hirschmann}}}, \bibinfo {author}
  {\bibfnamefont{L.}~\bibnamefont{{Lehner}}}, \bibinfo {author}
  {\bibfnamefont{S.~L.}\ \bibnamefont{{Liebling}}}, \bibinfo {author}
  {\bibfnamefont{P.~M.}\ \bibnamefont{{Motl}}},\ and\ \bibinfo {author}
  {\bibfnamefont{D.}~\bibnamefont{{Neilsen}}},\ }%
  \bibfield{journal}{%
  \Doi{10.1103/PhysRevD.80.024012}{\bibinfo {journal} {\prd}}\ }%
  \textbf{\bibinfo {volume} {80}},\ \bibinfo {pages} {024012} (\bibinfo {month}
  {Jul.}\ \bibinfo {year} {2009}),\
  \Eprint{http://arxiv.org/abs/0905.3390}{arXiv:0905.3390 [astro-ph.HE]}%
  \bibAnnoteFile{NoStop}{MegAFHLLMN09}%
\bibitem{AndLMN10}%
  \BibitemOpen
  \bibfield{author}{%
  \bibinfo {author} {\bibfnamefont{M.}~\bibnamefont{{Anderson}}}, \bibinfo
  {author} {\bibfnamefont{L.}~\bibnamefont{{Lehner}}}, \bibinfo {author}
  {\bibfnamefont{M.}~\bibnamefont{{Megevand}}},\ and\ \bibinfo {author}
  {\bibfnamefont{D.}~\bibnamefont{{Neilsen}}},\ }%
  \bibfield{journal}{%
  \Doi{10.1103/PhysRevD.81.044004}{\bibinfo {journal} {\prd}}\ }%
  \textbf{\bibinfo {volume} {81}},\ \bibinfo {pages} {044004} (\bibinfo {month}
  {Feb.}\ \bibinfo {year} {2010})%
  \bibAnnoteFile{NoStop}{AndLMN10}%
\bibitem{FarLS10}%
  \BibitemOpen
  \bibfield{author}{%
  \bibinfo {author} {\bibfnamefont{B.~D.}\ \bibnamefont{{Farris}}}, \bibinfo
  {author} {\bibfnamefont{Y.~T.}\ \bibnamefont{{Liu}}},\ and\ \bibinfo {author}
  {\bibfnamefont{S.~L.}\ \bibnamefont{{Shapiro}}},\ }%
  \bibfield{journal}{%
  \Doi{10.1103/PhysRevD.81.084008}{\bibinfo {journal} {\prd}}\ }%
  \textbf{\bibinfo {volume} {81}},\ \bibinfo {pages} {084008} (\bibinfo {month}
  {Apr.}\ \bibinfo {year} {2010}),\
  \Eprint{http://arxiv.org/abs/0912.2096}{arXiv:0912.2096 [astro-ph.HE]}%
  \bibAnnoteFile{NoStop}{FarLS10}%
\bibitem{BodHBLS10}%
  \BibitemOpen
  \bibfield{author}{%
  \bibinfo {author} {\bibfnamefont{T.}~\bibnamefont{{Bode}}}, \bibinfo {author}
  {\bibfnamefont{R.}~\bibnamefont{{Haas}}}, \bibinfo {author}
  {\bibfnamefont{T.}~\bibnamefont{{Bogdanovi{\'c}}}}, \bibinfo {author}
  {\bibfnamefont{P.}~\bibnamefont{{Laguna}}},\ and\ \bibinfo {author}
  {\bibfnamefont{D.}~\bibnamefont{{Shoemaker}}},\ }%
  \bibfield{journal}{%
  \Doi{10.1088/0004-637X/715/2/1117}{\bibinfo {journal} {\apj}}\ }%
  \textbf{\bibinfo {volume} {715}},\ \bibinfo {pages} {1117} (\bibinfo {month}
  {Jun.}\ \bibinfo {year} {2010})%
  \bibAnnoteFile{NoStop}{BodHBLS10}%
\bibitem{MostaPRLYP10}%
  \BibitemOpen
  \bibfield{author}{%
  \bibinfo {author} {\bibfnamefont{P.}~\bibnamefont{{M{\"o}sta}}}, \bibinfo
  {author} {\bibfnamefont{C.}~\bibnamefont{{Palenzuela}}}, \bibinfo {author}
  {\bibfnamefont{L.}~\bibnamefont{{Rezzolla}}}, \bibinfo {author}
  {\bibfnamefont{L.}~\bibnamefont{{Lehner}}}, \bibinfo {author}
  {\bibfnamefont{S.}~\bibnamefont{{Yoshida}}},\ and\ \bibinfo {author}
  {\bibfnamefont{D.}~\bibnamefont{{Pollney}}},\ }%
  \bibfield{journal}{%
  \Doi{10.1103/PhysRevD.81.064017}{\bibinfo {journal} {\prd}}\ }%
  \textbf{\bibinfo {volume} {81}},\ \bibinfo {eid} {064017} (\bibinfo {month}
  {Mar.}\ \bibinfo {year} {2010}),\
  \Eprint{http://arxiv.org/abs/0912.2330}{arXiv:0912.2330 [gr-qc]}%
  \bibAnnoteFile{NoStop}{MostaPRLYP10}%
\bibitem{PalenzuelaGLL10}%
  \BibitemOpen
  \bibfield{author}{%
  \bibinfo {author} {\bibfnamefont{C.}~\bibnamefont{{Palenzuela}}}, \bibinfo
  {author} {\bibfnamefont{T.}~\bibnamefont{{Garrett}}}, \bibinfo {author}
  {\bibfnamefont{L.}~\bibnamefont{{Lehner}}},\ and\ \bibinfo {author}
  {\bibfnamefont{S.~L.}\ \bibnamefont{{Liebling}}},\ }%
  \bibfield{journal}{%
  \Doi{10.1103/PhysRevD.82.044045}{\bibinfo {journal} {\prd}}\ }%
  \textbf{\bibinfo {volume} {82}},\ \bibinfo {pages} {044045} (\bibinfo {month}
  {Aug.}\ \bibinfo {year} {2010})%
  \bibAnnoteFile{NoStop}{PalenzuelaGLL10}%
\bibitem{ZanottiRDP10}%
  \BibitemOpen
  \bibfield{author}{%
  \bibinfo {author} {\bibfnamefont{O.}~\bibnamefont{{Zanotti}}}, \bibinfo
  {author} {\bibfnamefont{L.}~\bibnamefont{{Rezzolla}}}, \bibinfo {author}
  {\bibfnamefont{L.}~\bibnamefont{{Del Zanna}}},\ and\ \bibinfo {author}
  {\bibfnamefont{C.}~\bibnamefont{{Palenzuela}}},\ }%
  \bibfield{journal}{%
  \Doi{10.1051/0004-6361/201014969}{\bibinfo {journal} {\aap}}\ }%
  \textbf{\bibinfo {volume} {523}},\ \bibinfo {eid} {A8} (\bibinfo {month}
  {Nov.}\ \bibinfo {year} {2010}),\
  \Eprint{http://arxiv.org/abs/1002.4185}{arXiv:1002.4185 [astro-ph.HE]}%
  \bibAnnoteFile{NoStop}{ZanottiRDP10}%
\bibitem{FarLS11}%
  \BibitemOpen
  \bibfield{author}{%
  \bibinfo {author} {\bibfnamefont{B.~D.}\ \bibnamefont{{Farris}}}, \bibinfo
  {author} {\bibfnamefont{Y.~T.}\ \bibnamefont{{Liu}}},\ and\ \bibinfo {author}
  {\bibfnamefont{S.~L.}\ \bibnamefont{{Shapiro}}},\ }%
  \bibfield{journal}{%
  \Doi{10.1103/PhysRevD.84.024024}{\bibinfo {journal} {\prd}}\ }%
  \textbf{\bibinfo {volume} {84}},\ \bibinfo {eid} {024024} (\bibinfo {month}
  {Jul.}\ \bibinfo {year} {2011}),\
  \Eprint{http://arxiv.org/abs/1105.2821}{arXiv:1105.2821 [astro-ph.HE]}%
  \bibAnnoteFile{NoStop}{FarLS11}%
\bibitem{BodBHHLS12}%
  \BibitemOpen
  \bibfield{author}{%
  \bibinfo {author} {\bibfnamefont{T.}~\bibnamefont{{Bode}}}, \bibinfo {author}
  {\bibfnamefont{T.}~\bibnamefont{{Bogdanovi{\'c}}}}, \bibinfo {author}
  {\bibfnamefont{R.}~\bibnamefont{{Haas}}}, \bibinfo {author}
  {\bibfnamefont{J.}~\bibnamefont{{Healy}}}, \bibinfo {author}
  {\bibfnamefont{P.}~\bibnamefont{{Laguna}}},\ and\ \bibinfo {author}
  {\bibfnamefont{D.}~\bibnamefont{{Shoemaker}}},\ }%
  \bibfield{journal}{%
  \Doi{10.1088/0004-637X/744/1/45}{\bibinfo {journal} {\apj}}\ }%
  \textbf{\bibinfo {volume} {744}},\ \bibinfo {eid} {45} (\bibinfo {month}
  {Jan.}\ \bibinfo {year} {2012}),\
  \Eprint{http://arxiv.org/abs/1101.4684}{arXiv:1101.4684 [gr-qc]}%
  \bibAnnoteFile{NoStop}{BodBHHLS12}%
\bibitem{NobMNKCZY12}%
  \BibitemOpen
  \bibfield{author}{%
  \bibinfo {author} {\bibfnamefont{S.~C.}\ \bibnamefont{{Noble}}}, \bibinfo
  {author} {\bibfnamefont{B.~C.}\ \bibnamefont{{Mundim}}}, \bibinfo {author}
  {\bibfnamefont{H.}~\bibnamefont{{Nakano}}}, \bibinfo {author}
  {\bibfnamefont{J.~H.}\ \bibnamefont{{Krolik}}}, \bibinfo {author}
  {\bibfnamefont{M.}~\bibnamefont{{Campanelli}}}, \bibinfo {author}
  {\bibfnamefont{Y.}~\bibnamefont{{Zlochower}}},\ and\ \bibinfo {author}
  {\bibfnamefont{N.}~\bibnamefont{{Yunes}}},\ }%
  \bibfield{journal}{%
  \Doi{10.1088/0004-637X/755/1/51}{\bibinfo {journal} {\apj}}\ }%
  \textbf{\bibinfo {volume} {755}},\ \bibinfo {eid} {51} (\bibinfo {month}
  {Aug.}\ \bibinfo {year} {2012}),\
  \Eprint{http://arxiv.org/abs/1204.1073}{arXiv:1204.1073 [astro-ph.HE]}%
  \bibAnnoteFile{NoStop}{NobMNKCZY12}%
\bibitem{FarGPES12}%
  \BibitemOpen
  \bibfield{author}{%
  \bibinfo {author} {\bibfnamefont{B.~D.}\ \bibnamefont{{Farris}}}, \bibinfo
  {author} {\bibfnamefont{R.}~\bibnamefont{{Gold}}}, \bibinfo {author}
  {\bibfnamefont{V.}~\bibnamefont{{Paschalidis}}}, \bibinfo {author}
  {\bibfnamefont{Z.~B.}\ \bibnamefont{{Etienne}}},\ and\ \bibinfo {author}
  {\bibfnamefont{S.~L.}\ \bibnamefont{{Shapiro}}},\ }%
  \bibfield{journal}{%
  \Doi{10.1103/PhysRevLett.109.221102}{\bibinfo {journal} {Physical Review
  Letters}}\ }%
  \textbf{\bibinfo {volume} {109}},\ \bibinfo {eid} {221102} (\bibinfo {month}
  {Nov.}\ \bibinfo {year} {2012}),\
  \Eprint{http://arxiv.org/abs/1207.3354}{arXiv:1207.3354 [astro-ph.HE]}%
  \bibAnnoteFile{NoStop}{FarGPES12}%
\bibitem{Pri81}%
  \BibitemOpen
  \bibfield{author}{%
  \bibinfo {author} {\bibfnamefont{J.~E.}\ \bibnamefont{{Pringle}}},\ }%
  \bibfield{journal}{%
  \Doi{10.1146/annurev.aa.19.090181.001033}{\bibinfo {journal} {{Ann.\ Rev.\
  Astron.\ Astrophys.}}}\ }%
  \textbf{\bibinfo {volume} {19}},\ \bibinfo {pages} {137} (\bibinfo {year}
  {1981})%
  \bibAnnoteFile{NoStop}{Pri81}%
\bibitem{FraKR02}%
  \BibitemOpen
  \bibfield{author}{%
  \bibinfo {author} {\bibfnamefont{J.}~\bibnamefont{{Frank}}}, \bibinfo
  {author} {\bibfnamefont{A.}~\bibnamefont{{King}}},\ and\ \bibinfo {author}
  {\bibfnamefont{D.~J.}\ \bibnamefont{{Raine}}},\ }%
  \emph{\bibinfo {title} {{Accretion Power in Astrophysics}}}\ (\bibinfo
  {publisher} {Cambridge University Press, Cambridge},\ \bibinfo {year}
  {2002})%
  \bibAnnoteFile{NoStop}{FraKR02}%
\bibitem{GolT80}%
  \BibitemOpen
  \bibfield{author}{%
  \bibinfo {author} {\bibfnamefont{P.}~\bibnamefont{{Goldreich}}}\ and\
  \bibinfo {author} {\bibfnamefont{S.}~\bibnamefont{{Tremaine}}},\ }%
  \bibfield{journal}{%
  \Doi{10.1086/158356}{\bibinfo {journal} {\apj}}\ }%
  \textbf{\bibinfo {volume} {241}},\ \bibinfo {pages} {425} (\bibinfo {month}
  {Oct.}\ \bibinfo {year} {1980})%
  \bibAnnoteFile{NoStop}{GolT80}%
\bibitem{HouW84}%
  \BibitemOpen
  \bibfield{author}{%
  \bibinfo {author} {\bibfnamefont{K.}~\bibnamefont{{Hourigan}}}\ and\ \bibinfo
  {author} {\bibfnamefont{W.~R.}\ \bibnamefont{{Ward}}},\ }%
  \bibfield{journal}{%
  \Doi{10.1016/0019-1035(84)90136-2}{\bibinfo {journal} {Icarus}}\ }%
  \textbf{\bibinfo {volume} {60}},\ \bibinfo {pages} {29} (\bibinfo {month}
  {Oct.}\ \bibinfo {year} {1984})%
  \bibAnnoteFile{NoStop}{HouW84}%
\bibitem{LinP86}%
  \BibitemOpen
  \bibfield{author}{%
  \bibinfo {author} {\bibfnamefont{D.~N.~C.}\ \bibnamefont{{Lin}}}\ and\
  \bibinfo {author} {\bibfnamefont{J.}~\bibnamefont{{Papaloizou}}},\ }%
  \bibfield{journal}{%
  \Doi{10.1086/164653}{\bibinfo {journal} {\apj}}\ }%
  \textbf{\bibinfo {volume} {309}},\ \bibinfo {pages} {846} (\bibinfo {month}
  {Oct.}\ \bibinfo {year} {1986})%
  \bibAnnoteFile{NoStop}{LinP86}%
\bibitem{War97}%
  \BibitemOpen
  \bibfield{author}{%
  \bibinfo {author} {\bibfnamefont{W.~R.}\ \bibnamefont{{Ward}}},\ }%
  \bibfield{journal}{%
  \Doi{10.1006/icar.1996.5647}{\bibinfo {journal} {Icarus}}\ }%
  \textbf{\bibinfo {volume} {126}},\ \bibinfo {pages} {261} (\bibinfo {month}
  {Apr.}\ \bibinfo {year} {1997})%
  \bibAnnoteFile{NoStop}{War97}%
\bibitem{Cha08}%
  \BibitemOpen
  \bibfield{author}{%
  \bibinfo {author} {\bibfnamefont{P.}~\bibnamefont{{Chang}}},\ }%
  \bibfield{journal}{%
  \Doi{10.1086/590326}{\bibinfo {journal} {\apj}}\ }%
  \textbf{\bibinfo {volume} {684}},\ \bibinfo {pages} {236} (\bibinfo {month}
  {Sep.}\ \bibinfo {year} {2008}),\
  \Eprint{http://arxiv.org/abs/0801.2133}{arXiv:0801.2133}%
  \bibAnnoteFile{NoStop}{Cha08}%
\bibitem{GooR01}%
  \BibitemOpen
  \bibfield{author}{%
  \bibinfo {author} {\bibfnamefont{J.}~\bibnamefont{{Goodman}}}\ and\ \bibinfo
  {author} {\bibfnamefont{R.~R.}\ \bibnamefont{{Rafikov}}},\ }%
  \bibfield{journal}{%
  \Doi{10.1086/320572}{\bibinfo {journal} {\apj}}\ }%
  \textbf{\bibinfo {volume} {552}},\ \bibinfo {pages} {793} (\bibinfo {month}
  {May}\ \bibinfo {year} {2001}),\
  \Eprint{http://arxiv.org/abs/arXiv:astro-ph/0010576}{arXiv:astro-ph/0010576}%
  \bibAnnoteFile{NoStop}{GooR01}%
\bibitem{NovT73}%
  \BibitemOpen
  \bibfield{author}{%
  \bibinfo {author} {\bibfnamefont{I.~D.}\ \bibnamefont{{Novikov}}}\ and\
  \bibinfo {author} {\bibfnamefont{K.~S.}\ \bibnamefont{{Thorne}}},\ }%
  in\ \emph{\bibinfo {booktitle} {Black Holes, Les Houches}},\ \bibinfo
  {editor} {edited by\ \bibinfo {editor}
  {\bibfnamefont{C.}~\bibnamefont{{Dewitt}}}\ and\ \bibinfo {editor}
  {\bibfnamefont{B.}~\bibnamefont{{DeWitt}}}}\ (\bibinfo {publisher} {Gordon
  and Breach, New York},\ \bibinfo {year} {1973})\ pp.\ \bibinfo {pages}
  {343--450}%
  \bibAnnoteFile{NoStop}{NovT73}%
\bibitem{PagT74}%
  \BibitemOpen
  \bibfield{author}{%
  \bibinfo {author} {\bibfnamefont{D.~N.}\ \bibnamefont{{Page}}}\ and\ \bibinfo
  {author} {\bibfnamefont{K.~S.}\ \bibnamefont{{Thorne}}},\ }%
  \bibfield{journal}{%
  \Doi{10.1086/152990}{\bibinfo {journal} {\apj}}\ }%
  \textbf{\bibinfo {volume} {191}},\ \bibinfo {pages} {499} (\bibinfo {month}
  {Jul.}\ \bibinfo {year} {1974})%
  \bibAnnoteFile{NoStop}{PagT74}%
\bibitem{LigE75}%
  \BibitemOpen
  \bibfield{author}{%
  \bibinfo {author} {\bibfnamefont{D.~M.}\ \bibnamefont{{Eardley}}}\ and\
  \bibinfo {author} {\bibfnamefont{A.~P.}\ \bibnamefont{{Lightman}}},\ }%
  \bibfield{journal}{%
  \Doi{10.1086/153777}{\bibinfo {journal} {\apj}}\ }%
  \textbf{\bibinfo {volume} {200}},\ \bibinfo {pages} {187} (\bibinfo {month}
  {Aug.}\ \bibinfo {year} {1975})%
  \bibAnnoteFile{NoStop}{LigE75}%
\bibitem{Hir11a}%
  \BibitemOpen
  \bibfield{author}{%
  \bibinfo {author} {\bibfnamefont{C.~M.}\ \bibnamefont{{Hirata}}},\ }%
  \bibfield{journal}{%
  \Doi{10.1111/j.1365-2966.2011.18617.x}{\bibinfo {journal} {\mnras}}\ }%
  \textbf{\bibinfo {volume} {414}},\ \bibinfo {pages} {3198} (\bibinfo {month}
  {Jul.}\ \bibinfo {year} {2011}),\
  \Eprint{http://arxiv.org/abs/1010.0758}{arXiv:1010.0758 [astro-ph.HE]}%
  \bibAnnoteFile{NoStop}{Hir11a}%
\bibitem{Hir11b}%
  \BibitemOpen
  \bibfield{author}{%
  \bibinfo {author} {\bibfnamefont{C.~M.}\ \bibnamefont{{Hirata}}},\ }%
  \bibfield{journal}{%
  \Doi{10.1111/j.1365-2966.2011.18619.x}{\bibinfo {journal} {\mnras}}\ }%
  \textbf{\bibinfo {volume} {414}},\ \bibinfo {pages} {3212} (\bibinfo {month}
  {Jul.}\ \bibinfo {year} {2011}),\
  \Eprint{http://arxiv.org/abs/1010.0759}{arXiv:1010.0759 [astro-ph.HE]}%
  \bibAnnoteFile{NoStop}{Hir11b}%
\bibitem{ShaT83}%
  \BibitemOpen
  \bibfield{author}{%
  \bibinfo {author} {\bibfnamefont{S.~L.}\ \bibnamefont{{Shapiro}}}\ and\
  \bibinfo {author} {\bibfnamefont{S.~A.}\ \bibnamefont{{Teukolsky}}},\ }%
  \emph{\bibinfo {title} {{Black Holes, White Dwarfs, and Neutron Stars: The
  Physics of Compact Objects}}}\ (\bibinfo {publisher} {John Wiley, New York},\
  \bibinfo {year} {1983})%
  \bibAnnoteFile{NoStop}{ShaT83}%
\bibitem{BauS10}%
  \BibitemOpen
  \bibfield{author}{%
  \bibinfo {author} {\bibfnamefont{T.~W.}\ \bibnamefont{{Baumgarte}}}\ and\
  \bibinfo {author} {\bibfnamefont{S.~L.}\ \bibnamefont{{Shapiro}}},\ }%
  \emph{\bibinfo {title} {{Numerical Relativity: Solving Einstein's Equations
  on the Computer}}}\ (\bibinfo {publisher} {Cambridge University Press,
  Cambridge},\ \bibinfo {year} {2010})%
  \bibAnnoteFile{NoStop}{BauS10}%
\bibitem{PoiPV11}%
  \BibitemOpen
  \bibfield{author}{%
  \bibinfo {author} {\bibfnamefont{E.}~\bibnamefont{{Poisson}}}, \bibinfo
  {author} {\bibfnamefont{A.}~\bibnamefont{{Pound}}},\ and\ \bibinfo {author}
  {\bibfnamefont{I.}~\bibnamefont{{Vega}}},\ }%
  \bibfield{journal}{%
  \bibinfo {journal} {Living Reviews in Relativity}\ }%
  \textbf{\bibinfo {volume} {14}},\ \bibinfo {pages} {7} (\bibinfo {month}
  {Sep.}\ \bibinfo {year} {2011}),\
  \Eprint{http://arxiv.org/abs/1102.0529}{arXiv:1102.0529 [gr-qc]}%
  \bibAnnoteFile{NoStop}{PoiPV11}%
\bibitem{YunBHPBMT11}%
  \BibitemOpen
  \bibfield{author}{%
  \bibinfo {author} {\bibfnamefont{N.}~\bibnamefont{{Yunes}}}, \bibinfo
  {author} {\bibfnamefont{A.}~\bibnamefont{{Buonanno}}}, \bibinfo {author}
  {\bibfnamefont{S.~A.}\ \bibnamefont{{Hughes}}}, \bibinfo {author}
  {\bibfnamefont{Y.}~\bibnamefont{{Pan}}}, \bibinfo {author}
  {\bibfnamefont{E.}~\bibnamefont{{Barausse}}}, \bibinfo {author}
  {\bibfnamefont{M.~C.}\ \bibnamefont{{Miller}}},\ and\ \bibinfo {author}
  {\bibfnamefont{W.}~\bibnamefont{{Throwe}}},\ }%
  \bibfield{journal}{%
  \Doi{10.1103/PhysRevD.83.044044}{\bibinfo {journal} {\prd}}\ }%
  \textbf{\bibinfo {volume} {83}},\ \bibinfo {eid} {044044} (\bibinfo {month}
  {Feb.}\ \bibinfo {year} {2011}),\
  \Eprint{http://arxiv.org/abs/1009.6013}{arXiv:1009.6013 [gr-qc]}%
  \bibAnnoteFile{NoStop}{YunBHPBMT11}%
\bibitem{KulPSSNSZMDM11}%
  \BibitemOpen
  \bibfield{author}{%
  \bibinfo {author} {\bibfnamefont{A.~K.}\ \bibnamefont{{Kulkarni}}}, \bibinfo
  {author} {\bibfnamefont{R.~F.}\ \bibnamefont{{Penna}}}, \bibinfo {author}
  {\bibfnamefont{R.~V.}\ \bibnamefont{{Shcherbakov}}}, \bibinfo {author}
  {\bibfnamefont{J.~F.}\ \bibnamefont{{Steiner}}}, \bibinfo {author}
  {\bibfnamefont{R.}~\bibnamefont{{Narayan}}}, \bibinfo {author}
  {\bibfnamefont{A.}~\bibnamefont{{S{\"a} Dowski}}}, \bibinfo {author}
  {\bibfnamefont{Y.}~\bibnamefont{{Zhu}}}, \bibinfo {author}
  {\bibfnamefont{J.~E.}\ \bibnamefont{{McClintock}}}, \bibinfo {author}
  {\bibfnamefont{S.~W.}\ \bibnamefont{{Davis}}},\ and\ \bibinfo {author}
  {\bibfnamefont{J.~C.}\ \bibnamefont{{McKinney}}},\ }%
  \bibfield{journal}{%
  \Doi{10.1111/j.1365-2966.2011.18446.x}{\bibinfo {journal} {\mnras}}\ }%
  \textbf{\bibinfo {volume} {414}},\ \bibinfo {pages} {1183} (\bibinfo {month}
  {Jun.}\ \bibinfo {year} {2011}),\
  \Eprint{http://arxiv.org/abs/1102.0010}{arXiv:1102.0010 [astro-ph.HE]}%
  \bibAnnoteFile{NoStop}{KulPSSNSZMDM11}%
\bibitem{Sad09}%
  \BibitemOpen
  \bibfield{author}{%
  \bibinfo {author} {\bibfnamefont{A.}~\bibnamefont{{S{\c a}dowski}}},\ }%
  \bibfield{journal}{%
  \Doi{10.1088/0067-0049/183/2/171}{\bibinfo {journal} {\apjs}}\ }%
  \textbf{\bibinfo {volume} {183}},\ \bibinfo {pages} {171} (\bibinfo {month}
  {Aug.}\ \bibinfo {year} {2009}),\
  \Eprint{http://arxiv.org/abs/0906.0355}{arXiv:0906.0355 [astro-ph.HE]}%
  \bibAnnoteFile{NoStop}{Sad09}%
\bibitem{DolGML09}%
  \BibitemOpen
  \bibfield{author}{%
  \bibinfo {author} {\bibfnamefont{J.~C.}\ \bibnamefont{{Dolence}}}, \bibinfo
  {author} {\bibfnamefont{C.~F.}\ \bibnamefont{{Gammie}}}, \bibinfo {author}
  {\bibfnamefont{M.}~\bibnamefont{{Mo{\'s}cibrodzka}}},\ and\ \bibinfo {author}
  {\bibfnamefont{P.~K.}\ \bibnamefont{{Leung}}},\ }%
  \bibfield{journal}{%
  \Doi{10.1088/0067-0049/184/2/387}{\bibinfo {journal} {\apjs}}\ }%
  \textbf{\bibinfo {volume} {184}},\ \bibinfo {pages} {387} (\bibinfo {month}
  {Oct.}\ \bibinfo {year} {2009}),\
  \Eprint{http://arxiv.org/abs/0909.0708}{arXiv:0909.0708 [astro-ph.HE]}%
  \bibAnnoteFile{NoStop}{DolGML09}%
\bibitem{SchH11}%
  \BibitemOpen
  \bibfield{author}{%
  \bibinfo {author} {\bibfnamefont{R.~V.}\ \bibnamefont{{Shcherbakov}}}\ and\
  \bibinfo {author} {\bibfnamefont{L.}~\bibnamefont{{Huang}}},\ }%
  \bibfield{journal}{%
  \Doi{10.1111/j.1365-2966.2010.17502.x}{\bibinfo {journal} {\mnras}}\ }%
  \textbf{\bibinfo {volume} {410}},\ \bibinfo {pages} {1052} (\bibinfo {month}
  {Jan.}\ \bibinfo {year} {2011}),\
  \Eprint{http://arxiv.org/abs/1007.4831}{arXiv:1007.4831 [astro-ph.HE]}%
  \bibAnnoteFile{NoStop}{SchH11}%
\bibitem{VinPGP11}%
  \BibitemOpen
  \bibfield{author}{%
  \bibinfo {author} {\bibfnamefont{F.~H.}\ \bibnamefont{{Vincent}}}, \bibinfo
  {author} {\bibfnamefont{T.}~\bibnamefont{{Paumard}}}, \bibinfo {author}
  {\bibfnamefont{E.}~\bibnamefont{{Gourgoulhon}}},\ and\ \bibinfo {author}
  {\bibfnamefont{G.}~\bibnamefont{{Perrin}}},\ }%
  \bibfield{journal}{%
  \Doi{10.1088/0264-9381/28/22/225011}{\bibinfo {journal} {Classical and
  Quantum Gravity}}\ }%
  \textbf{\bibinfo {volume} {28}},\ \bibinfo {pages} {225011} (\bibinfo {month}
  {Nov.}\ \bibinfo {year} {2011}),\
  \Eprint{http://arxiv.org/abs/1109.4769}{arXiv:1109.4769 [gr-qc]}%
  \bibAnnoteFile{NoStop}{VinPGP11}%
\end{thebibliography}%
\end{document}